\documentclass[12pt]{article}
\usepackage{graphicx,amsfonts,amsmath,amssymb,mathrsfs}

\title{On the Distribution of the Wave
   Function for Systems in Thermal Equilibrium} 
\author{ 
Sheldon Goldstein\footnote{Departments of Mathematics and Physics,
     Hill Center, Rutgers, The State University of New  
     Jersey, 110 Frelinghuysen Road, Piscataway, NJ 08854-8019, USA.
     E-mail: oldstein@math.rutgers.edu},
Joel L. Lebowitz\footnote{Departments of Mathematics and Physics,
     Hill Center, Rutgers, The State University of New  
     Jersey, 110 Frelinghuysen Road, Piscataway, NJ 08854-8019, USA.
     E-mail: lebowitz@math.rutgers.edu},\\
Roderich Tumulka\footnote{Mathematisches Institut,
Eberhard-Karls-Universit\"at, Auf der Morgenstelle 10, 72076 T\"ubingen,
Germany.  E-mail: tumulka@everest.mathematik.uni-tuebingen.de}, and
Nino Zangh\`\i\footnote{Dipartimento di Fisica dell'Universit\`a di
    Genova and INFN sezione di Genova, Via Dodecaneso 33, 16146
    Genova, Italy.  E-mail: zanghi@ge.infn.it}
}
\date{August 12, 2005}

\newcommand{\Hilbert}{\mathscr{H}}
\newcommand{\conf}{\mathcal{Q}}
\newcommand{\tr}{\mathrm{tr}}
\renewcommand{\Re}{\mathrm{Re}}
\renewcommand{\Im}{\mathrm{Im}}
\newcommand{\EEE}{\mathbb{E}}
\newcommand{\PPP}{\mathbb{P}}
\newcommand{\RRR}{\mathbb{R}}
\newcommand{\CCC}{\mathbb{C}}
\newcommand{\QQQ}{\mathbb{Q}}

\newcommand{\sphere}{\mathscr{S}} 
\renewcommand{\sp}[2]{\langle #1|#2 \rangle}
\newcommand{\spec}{\mathrm{spec}} 
\newcommand{\support}{\mathrm{support}}
\newcommand{\R}{\mathscr{R}}
\newcommand{\G}[1]{G(#1)}
\newcommand{\GA}[1]{GA(#1)}
\newcommand{\GAP}[1]{GAP(#1)}
\newcommand{\EIG}[1]{EIG(#1)}
\newcommand{\red}{\mathrm{red}}
\newcommand{\proj}{P}

\newcommand{\onb}{$\{ |q_2\rangle\}$}

\newtheorem{lemma}{Lemma}
\newenvironment{proof}[1]{\medskip\noindent\textit{Proof#1.}}
	       {$\hfill\square$\medskip}

\newcommand{\z}[1]{{#1}}
\newcommand{\zz}[1]{{#1}}

\begin{document}
\maketitle
\begin{abstract}
  For a quantum system, a density matrix $\rho$ that is not pure can arise,
  via averaging, from a distribution $\mu$ of its wave function, a
  normalized vector belonging to its Hilbert space $\Hilbert$. While $\rho$
  itself does not determine a unique $\mu$, additional facts, such as that
  the system has come to thermal equilibrium, might. It is thus not
  unreasonable to ask, which $\mu$, if any, corresponds to a given
  thermodynamic ensemble?  To answer this question we construct, for any
  given density matrix $\rho$, a natural measure on the unit sphere in
  $\Hilbert$, denoted $\GAP{\rho}$. We do this using a suitable projection
  of the Gaussian measure on $\Hilbert$ with covariance $\rho$. We
  establish some nice properties of $\GAP{\rho}$ and show that this measure
  arises naturally when considering macroscopic systems. In particular, we
  argue that it is the most appropriate choice for systems in thermal
  equilibrium, described by the canonical ensemble density matrix
  $\rho_\beta = (1/Z) \exp(- \beta H)$. $\GAP{\rho}$ may also be relevant
  to  quantum chaos and to the stochastic evolution of open
  quantum systems, where distributions on $\Hilbert$ are often used.

%

  \medskip

\end{abstract}

\section{Introduction}

In classical mechanics, ensembles, such as the microcanonical and canonical
ensembles, are represented by probability distributions on the phase space.
In quantum mechanics, ensembles are usually represented by density
matrices.  \z{It is natural to regard} these density matrices as arising
from probability distributions on the (normalized) wave functions
associated with the thermodynamical ensembles, so that members of the
ensemble are represented by a random state vector.  There are, however, as
is well known, many probability distributions which give rise to the same
density matrix, and thus to the same predictions for experimental outcomes
\cite[sec.~IV.3]{Neumann}.\footnote{\z{This empirical equivalence should not
too hastily be regarded as implying physical equivalence. Consider, for
example, the two Schr\"odinger's cat states $\Psi_{\pm}=
(\Psi_{\text{alive}} \pm \Psi_{\text{dead}})/\sqrt2$. The measure that
gives equal weight to these two states corresponds to the same density
matrix as the one giving equal weight to $\Psi_{\text{alive}}$ and
$\Psi_{\text{dead}}$. However the physical situation corresponding to the
former measure, a mixture of two grotesque superpositions, seems
dramatically different from the one corresponding to the latter, a routine
mixture. It is thus not easy to regard these two measures as physically
equivalent.}}  Moreover, as emphasized by Landau and Lifshitz \cite[sec.
I.5]{LL59}, the energy levels for macroscopic systems are so closely spaced
(exponentially small in the number of particles in the system) that ``the
concept of stationary states [energy eigenfunctions] becomes in a certain
sense unrealistic'' because of the difficulty of preparing a system with
such a sharp energy and keeping it isolated. Landau and Lifshitz are
therefore wary of, and warn against, regarding the density matrix for such
a system as arising solely from our lack of knowledge about the wave
function of the system.  We shall argue, however, that despite these
caveats such distributions can be both useful and physically meaningful.
In particular we describe here a novel probability distribution, to be
associated with any thermal ensemble such as the canonical ensemble.

\z{While probability distributions on wave functions are natural objects of
study in many contexts,} \zz{from quantum chaos \cite{berry,sred,urbina} to
open quantum systems \cite{BP02},} \z{our main} motivation for considering
\z{them} is to exploit the analogy between classical and quantum
statistical mechanics \cite{schr,schrbook,Bloch,joel2,joel3,joel1}.  This
analogy suggests that some relevant classical reasonings can be transferred
to quantum mechanics by formally replacing the classical phase space by the
unit sphere $\sphere (\Hilbert)$ of the quantum system's Hilbert space
$\Hilbert$.  In particular, with a natural measure $\mu(d \psi)$ on
$\sphere(\Hilbert)$ one \z{can utilize the notion of typicality, i.e.,
consider properties of a system common to ``almost all'' members of an
ensemble. This is a notion frequently used in equilibrium statistical
mechanics, as in, e.g., Boltzmann's recognition that typical phase points
on the energy surface of a macroscopic system are such that the empirical
distribution of velocities is approximately Maxwellian. Once one has such a
\z{measure} for quantum systems, one} could attempt an analysis of the
second law of thermodynamics in quantum mechanics along the lines of
Boltzmann's analysis of the second law in classical mechanics, involving an
argument to the effect that the behavior described in the second law (such
as entropy increase) occurs for typical states of an isolated macroscopic
system, i.e.\ for the overwhelming majority of points on
$\sphere(\Hilbert)$ with respect to $\mu(d\psi)$.

\z{Probability} distributions on wave functions of a composite system, with
Hilbert space $\Hilbert$, have \z{in fact} been used to establish the
typical properties of the reduced density matrix of a subsystem arising
from the wave function of the composite. For example, Page \cite{Pag93}
considers the uniform distribution on $\sphere(\Hilbert)$ for a
finite-dimensional Hilbert space $\Hilbert$, in terms of which he shows
that the von Neumann entropy of the reduced density matrix is typically
nearly maximal under appropriate conditions on the dimensions of the
relevant Hilbert spaces.

Given a probability distribution $\mu$ on the unit sphere
$\sphere(\Hilbert)$ of the Hilbert space $\Hilbert$ there is always an
associated density matrix $\rho_\mu$ \cite{Neumann}: it is the density
matrix of the mixture, or the statistical ensemble of systems, defined by
the distribution $\mu$, given by
\begin{equation}\label{rhomupsi}
  \rho_\mu = \int\limits_{\sphere(\Hilbert)} \!\!\! \mu(d\psi) \,
  |\psi\rangle \langle\psi| \,.
\end{equation}
For any projection operator $P$, $\tr\,(\rho_\mu P)$ is the
probability of obtaining in an experiment a result corresponding to
$P$ for a system with a $\mu$-distributed wave function.  It is
evident from \eqref{rhomupsi} that $\rho_\mu$ is  the second
moment, or covariance matrix, of $\mu$, provided $\mu$ has mean 0
(which may, and will, be assumed without loss of generality since
$\psi$ and $-\psi$ are equivalent physically).

While a probability measure $\mu$ on $\sphere(\Hilbert)$ determines a
unique density matrix $\rho$ on $\Hilbert$ via \eqref{rhomupsi}, the
converse is not true: the association $\mu \mapsto \rho_\mu$ given by
\eqref{rhomupsi} is many-to-one.\footnote{For example, in a $k$-dimensional
Hilbert space the uniform probability distribution
$u=u_{\sphere(\Hilbert)}$ over the unit sphere has density matrix $\rho_u =
\tfrac{1}{k}I$ with $I$ the identity operator on $\Hilbert$; at the same
time, for every orthonormal basis of $\Hilbert$ the uniform distribution
over the basis (which is a measure concentrated on just $k$ points) has the
same density matrix, $\rho = \tfrac{1}{k} I$. An exceptional case is the
density matrix corresponding to a pure state, $\rho = |\psi \rangle \langle
\psi|$, as the measure $\mu$ with this density matrix is almost unique: it
must be concentrated on the ray through $\psi$, and thus the only
non-uniqueness corresponds to the distribution of the phase.}  There is
furthermore no unique ``physically correct'' choice of $\mu$ for a given
$\rho$ since for any $\mu$ corresponding to $\rho$ one could, in principle,
prepare an ensemble of systems with wave functions distributed according to
this $\mu$.  However, \z{while $\rho$ itself need not determine a unique
probability measure, additional facts about a system, such as that it has
come to} thermal equilibrium, \z{might}. \z{It is thus} not unreasonable to
ask: which measure on $\sphere(\Hilbert)$ corresponds to a \z{given}
thermodynamic ensemble?

Let us start with the \emph{microcanonical} ensemble, corresponding to the
energy interval $[E, E+\delta]$, where $\delta$ is small on the macroscopic
scale but large enough for the interval to contain many eigenvalues. To
this there is associated the spectral subspace $\Hilbert_{E,\delta}$, the
span of the eigenstates $|n\rangle$ of the Hamiltonian $H$ corresponding to
eigenvalues $E_n$ between $E$ and $E + \delta$. Since $\Hilbert_{E,\delta}$
is finite dimensional, one can form the \emph{microcanonical density
  matrix}
\begin{equation}\label{microcanrho}
  \rho_{E,\delta} = (\dim \Hilbert_{E,\delta})^{-1} P_{\Hilbert_{E,\delta}}
\end{equation}
with $P_{\Hilbert_{E,\delta}} = 1_{[E,E+\delta]}(H)$ the projection to
$\Hilbert_{E,\delta}$. This density matrix is diagonal in the energy
representation and gives equal weight to all energy eigenstates in the 
interval $[E, E+\delta]$. 

But what is the corresponding \emph{microcanonical measure}? The most
plausible answer, given long ago by Schr\"odinger
\cite{schr,schrbook} and Bloch \cite{Bloch}, is the
(normalized) uniform measure $u_{E,\delta} = u_{\sphere
  (\Hilbert_{E,\delta})}$ on the unit sphere in this subspace.
$\rho_{E,\delta}$ is associated with $u_{E,\delta}$ via \eqref{rhomupsi}.

Note that a
wave function $\Psi$ chosen at random from this distribution is almost
certainly a nontrivial superposition of the eigenstates $|n
\rangle$ with random coefficients $\sp{n}{\Psi}$ that are identically
distributed, but not independent.  The measure $u_{E,\delta}$ is clearly
stationary, i.e., invariant under the unitary time evolution generated by
$H$, and it is as spread out as it could be over the set
$\sphere(\Hilbert_{E,\delta})$ of allowed wave functions.  This measure
provides us with a notion of a ``typical wave function'' from
$\Hilbert_{E,\delta}$ which is very different from the one arising from the
measure $\mu_{E,\delta}$ that, when $H$ is nondegenerate, gives equal
probability $(\dim
\Hilbert_{E,\delta})^{-1}$ to every eigenstate $|n \rangle$ with
eigenvalue $E_n \in [E,E+\delta]$.  The measure $\mu_{E,\delta}$, which is
concentrated on these eigenstates, is, however, less robust to small
perturbations in $H$ than is the smoother measure $u_{E,\delta}$.

Our proposal for the canonical ensemble is in the spirit of the uniform
microcanonical measure $u_{E,\delta}$  and reduces to it in the appropriate
cases. It is based on a mathematically
natural family of probability measures $\mu$ on $\sphere(\Hilbert)$.
For every density matrix $\rho$ on $\Hilbert$, there is a unique
member $\mu$ of this family, satisfying \eqref{rhomupsi} for $\rho_\mu =
\rho$, namely the \emph{Gaussian adjusted projected measure}
$\GAP{\rho}$, constructed roughly as follows: Eq.~\eqref{rhomupsi}
(i.e., the fact that $\rho_\mu$ is the covariance of $\mu$) suggests
that we start by considering the Gaussian measure $\G{\rho}$ with
covariance $\rho$ (and mean 0), which could, in finitely many
dimensions, be expressed by $\G{\rho}(d\psi) \propto \exp (- \sp
{\psi} {\rho^{-1}| \psi}) \, d\psi$ (where $d\psi$ is the obvious
Lebesgue measure on $\Hilbert$).\footnote{\zz{Berry \cite{berry} has
conjectured, and for some cases proven, that such measures describe
interesting universal properties of chaotic energy eigenfunctions in the
semiclassical regime, see also \cite{sred,urbina}. It is perhaps worth
considering the possibility that the GAP measures described here provide
somewhat better candidates for this purpose.}} This is not adequate, however, since
the measure that we seek must live on the sphere $\sphere (\Hilbert)$
whereas $\G{\rho}$ is spread out over all of $\Hilbert$.  We thus
adjust and then project $\G{\rho}$ to $\sphere (\Hilbert)$, in the
manner described in Section~\ref{sec:defmeasure}, in order to obtain
the measure $\GAP{\rho}$, having the prescribed covariance $\rho$ as
well as other desirable properties.

It is our contention that \textit{a quantum system in thermal
equilibrium at inverse temperature $\beta$ should be described by a
random state vector whose distribution is given by the measure
$\GAP{\rho_\beta}$ associated with the density matrix for the
canonical ensemble,}
\begin{equation}\label{rhobetaH}
  \rho_\beta = \rho_{\Hilbert,H,\beta} = \frac{1}{Z} \exp(-\beta H)
  \text{ with } Z:=\tr\, \exp(-\beta H).
\end{equation}

In order to convey the significance of $\GAP{\rho}$ as well as the
plausibility of our proposal that $\GAP{\rho_\beta}$ describes thermal
equilibrium, we recall that a system described by a canonical ensemble is
usually regarded as a subsystem of a larger system. It is therefore
important to consider the notion of the distribution of the wave function
of a subsystem.  Consider a composite system in a pure state $\psi \in
\Hilbert_1 \otimes \Hilbert_2$, and ask what might be meant by the wave
function of the subsystem with Hilbert space $\Hilbert_1$. For this we
propose the following.  Let $\{|q_2\rangle\}$ be a (generalized)
orthonormal basis of $\Hilbert_2$ (playing the role, say, of the eigenbasis
of the position representation).  \z{For each choice of $|q_2\rangle$, the
(partial) scalar product $\sp{q_2} {\psi}$, taken in $\Hilbert_2$, is a
vector belonging to $\Hilbert_1$. Regarding $|q_2\rangle$ as random, we
are led to consider} the random vector $\Psi_1\in \Hilbert_1$ given by
\begin{equation}\label{Psi1def}
  \Psi_1 = \mathcal{N} \: \sp{Q_2} {\psi}
\end{equation}
where $\mathcal{N}= \mathcal{N} (\psi,Q_2) = \bigl\| \sp{Q_2}{\psi}
\bigr\|^{-1}$ is the normalizing factor and $|Q_2\rangle$ is a random
element of the basis $\{|q_2\rangle\}$, chosen with the quantum
distribution
\begin{equation}\label{marg}
  \PPP(Q_2 = q_2) = \bigl\| \sp{q_2} {\psi} \bigr\|^2.
\end{equation}
\z{We refer to $\Psi_1$ as  the \emph{conditional wave function}
\cite{DGZ} of system 1.}
Note that $\Psi_1$ becomes doubly random when we start with a random
wave function in $\Hilbert_1 \otimes \Hilbert_2$ instead of a fixed
one.

The distribution of $\Psi_1$ corresponding to (\ref {Psi1def}) and
(\ref{marg}) is given by the probability measure on $\sphere
(\Hilbert_1)$,
\begin{equation}\label{mu1}
  \mu_1(d\psi_1)= \PPP(\Psi_1 \in d\psi_1) = \sum_{q_2} \bigl\|
  \sp{q_2} {\psi} \bigr\|^2 \, \delta\bigl( \psi_1 - \mathcal{N}
  (\psi, q_2) \, \sp{q_2} {\psi} \bigr) \, d\psi_1 \,,
\end{equation}
where $\delta(\psi - \phi) \, d\psi$ denotes the ``delta'' measure
concentrated at $\phi$. While the density matrix $\rho_{\mu_1}$ associated
with ${\mu_1}$ always equals the reduced density matrix $\rho_1^\red$ of
system 1, given by
\begin{equation}\label{rhoreddef}
  \rho_1^\red = \tr_2 |\psi \rangle \langle \psi | = \sum_{q_2}
  \sp{q_2}{\psi} \sp{\psi}{q_2} \,,
\end{equation}
the measure $\mu_1$ itself usually depends on the choice of the basis $\{|
q_2 \rangle\}$.  It turns out, nevertheless, as we point out in
Section~\ref{sec:typicality}, that $\mu_1(d\psi_1)$ is a universal function
of $\rho_1^\red$ in the special case that system 2 is large and $\psi$ is
typical (with respect to the uniform distribution on all wave functions
with the same reduced density matrix), namely $\GAP{\rho_1^\red}$. Thus
$\GAP{\rho}$ has a distinguished, universal status among all probability
measures on $\sphere(\Hilbert)$ with density matrix $\rho$.

To \z{further} support our claim that $\GAP{\rho_\beta}$ is the right
measure for $\rho_\beta$, we shall regard, as is usually done, the system
described by $\rho_\beta$ as coupled to a (very large) heat bath.  The
interaction between the heat bath and the system is assumed to be (in some
suitable sense) negligible. We will argue that if the wave function $\psi$
of the combined ``system plus bath'' has microcanonical distribution
$u_{E,\delta}$, then the distribution of the conditional wave function of
the (small) system is approximately $\GAP{\rho_{\beta}}$; see
Section~\ref{sec:hb1}.

Indeed, a stronger statement is  true.  As we argue in
Section~\ref{sec:hb2}, even for a typical
\emph{fixed} microcanonical wave function $\psi$ of the composite, i.e.,
one typical for $u_{E,\delta}$, the conditional wave function of the
system, defined in
\eqref{Psi1def}, is then approximately $\GAP{\rho_\beta}$-distributed, for
a typical basis $\{|q_2\rangle\}$. This is related to the fact that for a
typical microcanonical wave function $\psi$ of the composite the reduced
density matrix for the system is approximately $\rho_\beta$
\cite{thermo4, schrbook}. Note that the analogous statement in
classical mechanics would be wrong: for a fixed phase point $\xi$ of the
composite, be it typical or atypical, the phase point of the system could
never be random, but rather would merely be the part of $\xi$ belonging to
the system.

The remainder of this paper is organized as follows. In
Section~\ref{sec:defmeasure} we define the measure $\GAP{\rho}$ and obtain
several ways of writing it.  In Section~\ref{sec:prop} we describe \zz{some
natural} mathematical properties of these measures, \zz{and suggest that
these properties uniquely characterize the measures}. In
Section~\ref{sec:hb1} we argue that $\GAP{\rho_\beta}$ represents the
canonical ensemble.  In Section~\ref{sec:Typicality} we outline the proof
that $\GAP{\rho}$ is the distribution of the conditional wave function for
\emph{most} wave functions in $\Hilbert_1 \otimes \Hilbert_2$ with reduced
density matrix $\rho$ if system 2 is large, and show that
$\GAP{\rho_\beta}$ is the typical distribution of the conditional wave
function arising from a fixed microcanonical wave function of a \zz{system
in contact with a} heat bath.  In Section~\ref{sec:rem} we discuss other
measures that have been or might be considered as the thermal equilibrium
distribution of the wave function.  Finally, in Section~\ref{sec:two} we
compute explicitly the distribution of the coefficients of a
$\GAP{\rho_\beta}$-distributed state vector in the simplest possible
example, the two-level system.

\section{Definition of $\GAP{\rho}$}
\label{sec:defmeasure}

In this section, we define, for any given density matrix $\rho$ \z{on a
(separable) Hilbert space $\Hilbert,$} the Gaussian adjusted projected
measure $\GAP{\rho}$ on $\sphere (\Hilbert)$.  This definition makes use of
two auxiliary measures, $\G{\rho}$ and $\GA{\rho}$, defined as follows.

$\G{\rho}$ is the Gaussian measure on $\Hilbert$ with covariance
matrix $\rho$ (and mean 0).  More explicitly, let $\{|n\rangle\}$ be
an orthonormal basis of eigenvectors of $\rho$ and $p_n$ the
corresponding eigenvalues,
\begin{equation}
  \rho = \sum_n p_n \, |n \rangle \langle n|.
\end{equation}
Such a basis exists because $\rho$ has finite trace.  Let $Z_n$ be a
sequence of independent complex-valued random variables having a
(rotationally symmetric) Gaussian distribution in $\CCC$ with mean $0$
and variance
\begin{equation}\label{variance}
  \EEE |Z_n|^2 = p_n
\end{equation} 
(where $\EEE$ means expectation), i.e., $\Re \, Z_n$ and $\Im \, Z_n$
are independent real Gaussian variables with mean zero and variance
$p_n/2$.  We define $\G{\rho}$ to be the distribution of the random
vector
\begin{equation}\label{psiGdef}
  \Psi^G := \sum_n Z_n |n\rangle \,.
\end{equation}
Note that $\Psi^G$ is not normalized, i.e., it does not lie in
$\sphere (\Hilbert)$.  In order that $\Psi^G$ lie in $\Hilbert$ at
all, we need that the sequence $Z_n$ be square-summable, $\sum_n
|Z_n|^2 < \infty$.  That this is almost surely the case follows from the
fact that $\EEE \sum_n |Z_n|^2$ is finite. In fact, 
\begin{equation}\label{Esumvariances}
  \EEE \sum_n |Z_n|^2 = \sum_n \EEE |Z_n|^2 = \sum_n p_n = \tr\,
  \rho = 1.
\end{equation}
More generally, we observe that for any measure $\mu$ on $\Hilbert$
with (mean 0 and) covariance given by the trace class operator $C$,
\[
  \int_{\Hilbert} \mu(d\psi) \, |\psi \rangle \langle \psi| = C\,,
\]
we have that, for a random vector $\Psi$ with distribution $\mu$,
$\EEE \| \Psi\|^2 = \tr\, C$.

It also follows that $\Psi^G$ almost surely lies in the positive
spectral subspace of $\rho$, the closed subspace spanned by those
$|n\rangle$ with $p_n \neq 0$, or, equivalently, the orthogonal
complement of the kernel of $\rho$; we shall call this subspace
$\support(\rho)$.  Note further that, since $\G{\rho}$ is the Gaussian
measure with covariance $\rho$, it does not depend (in the case of
degenerate $\rho$) on the choice of the basis $\{|n\rangle \}$ among
the eigenbases of $\rho$, but only on $\rho$.

\z{Since we want a measure on $\sphere (\Hilbert)$ while $\G{\rho}$ is not
concentrated on $\sphere (\Hilbert)$ but rather is spread out, it would be
natural to project $\G{\rho}$ to $\sphere (\Hilbert)$.  However, since
projecting to $\sphere (\Hilbert)$ changes the covariance of a measure, as
we will point out in detail in Section \ref{sec:densitymatrix}, we
introduce an adjustment factor that exactly compensates for the change of
covariance due to projection.} We \z{thus} define the adjusted Gaussian measure
$\GA{\rho}$ on $\Hilbert$ by
\begin{equation}\label{muNdef}
  \GA{\rho} (d\psi) = \|\psi\|^2 \; \G{\rho} (d\psi).
\end{equation}
Since $\EEE \|\Psi^G\|^2 =1$ by \eqref{Esumvariances}, $\GA{\rho}$ is
a probability measure.  

Let $\Psi^{GA}$ be a $\GA{\rho}$-distributed random vector.  We define
$\GAP{\rho}$ to be  the distribution of 
\begin{equation}\label{mudef}
    \Psi^{GAP} :=
  \frac{\Psi^{GA}}{\|\Psi^{GA}\|} = \proj(\Psi^{GA})
\end{equation}
with $\proj$ the projection to the unit sphere (i.e., the
normalization of a vector),
\begin{equation}\label{projdef}
  \proj: \Hilbert \setminus \{0\} \to \sphere(\Hilbert)\,, \quad
  \proj(\psi) = \|\psi\|^{-1} \psi.
\end{equation}
Putting \eqref{mudef} differently, for a subset $B \subseteq
\sphere(\Hilbert)$,
\begin{equation}\label{muB}
 \GAP{\rho}(B) = \GA{\rho} (\RRR^+ B) = \int\limits_{\RRR^+ B}
 \G{\rho}(d\psi) \; \|\psi\|^2
\end{equation}
where $\RRR^+B$ denotes the cone through $B$. More succinctly,
\begin{equation}\label{GAPprojGA}
  \GAP{\rho} = \proj_* \bigl( \GA{\rho} \bigr) =
  \GA{\rho} \circ \proj^{-1}\,.
\end{equation}
where $\proj_*$ denotes the action of $\proj$ on measures. 

More generally, one can define for any measure $\mu$ on $\Hilbert$ the
``adjust-and-project'' procedure: let $A(\mu)$ be the adjusted measure
$A(\mu)(d\psi) = \|\psi\|^2 \, \mu(d\psi)$; then the
adjusted-and-projected measure is $\proj_* \bigl( A(\mu) \bigr) =
A(\mu) \circ \proj^{-1}$, thus defining a mapping $\proj_* \circ A$
from the measures on $\Hilbert$ with $\int \mu(d\psi) \, \|\psi\|^2 =
1$ to the probability measures on $\sphere(\Hilbert)$. We then have
that $\GAP{\rho} = \proj_* \bigl( A(\G{\rho}) \bigr)$.

We remark that $\Psi^{GAP}$, too, lies in $\support(\rho)$ almost
surely, and that $\proj(\Psi^G)$ does \emph{not} have distribution
$\GAP{\rho}$---nor covariance $\rho$ (see
Sect.~\ref{sec:densitymatrix}).

We can be more explicit in the case that $\rho$ has finite rank $k = \dim
\, \support(\rho)$, e.g.\ for finite-dimensional $\Hilbert$: then
there exists a Lebesgue volume measure $\lambda$ on $\support(\rho) =
\CCC^k$, and we can specify the densities of $\G{\rho}$ and
$\GA{\rho}$,
\begin{subequations}\label{muNdensity}
\begin{align}
  \frac{d\G{\rho}}{d\lambda}(\psi) &= \frac{1}{\pi^k \,
  \det \rho_+} \exp(-\langle \psi |\rho^{-1}_+| \psi \rangle),\\
  \frac{d\GA{\rho}}{d\lambda}(\psi) &= \frac{ \|\psi\|^2}{\pi^k \,
  \det \rho_+} \exp(-\langle \psi |\rho^{-1}_+| \psi \rangle),
\end{align}
\end{subequations}
with $\rho_+$ the restriction of $\rho$ to $\support(\rho)$.
Similarly, we can express $\GAP{\rho}$ relative to the
$(2k-1)$--dimensional surface measure $u$ on $\sphere(\support(\rho))$,
\begin{subequations}\label{mudensity}
\begin{align}
  \frac{d\GAP{\rho}}{du}(\psi) &= \frac{1}{\pi^k \, \det \rho_+}
  \int\limits_0^\infty dr \, r^{2k-1} \, r^2 \exp(-r^2 \langle \psi |
  \rho^{-1}_+ | \psi \rangle) = \\ &= \frac{k!}{2\pi^{k} \, \det
  \rho_+} \, \langle \psi | \rho^{-1}_+ | \psi \rangle^{-k-1}
  \,.\label{mupowerlaw}
\end{align}
\end{subequations}

We note that
\begin{equation}\label{microGAP}
  \GAP{\rho_{E,\delta}} = u_{E,\delta}\,,
\end{equation}
where $\rho_{E\delta}$ is the microcanonical density matrix given in
\eqref{microcanrho} and $u_{E\delta}$ is the microcanonical measure.

\section{Properties of $\GAP{\rho}$} 
\label{sec:prop}

In this section we prove the following properties of $\GAP{\rho}$:

\bigskip

\noindent{\bf Property 1}\textit{   The density matrix
  associated with $\GAP{\rho}$ in the sense of \eqref{rhomupsi} is $\rho$,
  i.e., $\rho_{\GAP{\rho}} = \rho$.
}

\bigskip

\noindent{\bf Property 2}\textit{    The association $\rho \mapsto
  \GAP{\rho}$ is \emph{covariant}: For any 
  unitary operator $U$ on $\Hilbert$, 
  \begin{equation}\label{covariance}
    U_* \GAP{\rho} = \GAP{U\rho U^*}
  \end{equation}
  where $U^* = U^{-1}$ is the adjoint of $U$ and $U_*$ is the action of $U$ on
  measures, $U_* \mu = \mu \circ U^{-1}$.  In particular, $\GAP{\rho}$ is
  stationary under any unitary evolution that preserves $\rho$.
}


\bigskip

\noindent{\bf Property 3}\textit{    If $\Psi \in \Hilbert_1 \otimes \Hilbert_2$ has distribution
  $\GAP{\rho_1 \otimes \rho_2}$ then, for any basis $\{|q_2\rangle\}$
  of $\Hilbert_2$, the conditional wave function $\Psi_1$ has
  distribution $\GAP{\rho_1}$. (``GAP of a product density matrix has
  GAP marginal.'')
}
\medskip

We will refer to the property expressed in Property~3 by
saying that the family of GAP measures is \emph{hereditary}.  We note
that when $\Psi \in \Hilbert_1 \otimes \Hilbert_2$ has distribution
$\GAP{\rho}$ and $\rho$ is not a tensor product, the distribution of
$\Psi_1$ need not be $\GAP{\rho_1^\red}$ (as we will show after the
proof of Property~3).

Before establishing these properties let us formulate what they say about
our candidate $\GAP{\rho_\beta}$ for the canonical distribution.  As a
consequence of Property~1, the density matrix arising from $\mu =
\GAP{\rho_\beta}$ in the sense of \eqref{rhomupsi} is the density matrix
$\rho_{\beta}$. As a consequence of Property~2, $\GAP{\rho_\beta}$
is stationary, i.e., invariant under the unitary time evolution generated
by $H$. As a consequence of Property~3, if $\Psi \in \Hilbert =
\Hilbert_1 \otimes \Hilbert_2$ has distribution $\GAP{\rho_{\Hilbert,
    H, \beta}}$ and systems 1 and 2 are decoupled, $H = H_1 \otimes I_2 +
I_1 \otimes H_2$, where $I_i$ is the identity on $\Hilbert_i$, then the
conditional wave function $\Psi_1$ of system $1$ has a distribution (in
$\Hilbert_1$) of the same kind with the same inverse temperature $\beta$,
namely $\GAP{\rho_{\Hilbert_1, H_1, \beta}}$.  This fits well with our
claim that $\GAP{\rho_\beta}$ is the thermal equilibrium distribution since
one would expect that if a system is in thermal equilibrium at inverse
temperature $\beta$ then so are its subsystems.

\z{We conjecture that the family of GAP measures is the only family of
measures satisfying Properties~1--3. This conjecture is formulated in
detail, and established for suitably continuous families of measures, in
Section \ref{sec:uniqueness}.}

The following lemma, proven in Section~\ref{gmgm}, is convenient for
showing that a random wave function is GAP-distributed:

\begin{lemma}\label{GaussGAP}
  Let $\Omega$ be a measurable space, $\mu$ a probability measure on
  $\Omega$, and $\Psi: \Omega \to \Hilbert$ a Hilbert-space-valued
  function.  If $\Psi(\omega)$ is $\G{\rho}$-distributed with respect
  to $\mu(d\omega)$, then $\Psi(\omega)/\|\Psi(\omega)\|$ is
  $\GAP{\rho}$-distributed with respect to $\|\Psi(\omega)\|^2
  \mu(d\omega)$.
\end{lemma}

\subsection{The Density Matrix} \label{sec:densitymatrix}

In this subsection we establish Property~1. We then add a
remark on the covariance matrix.

\begin{proof}{ of Property~1} {}From \eqref{rhomupsi} we find
that
\begin{align}\nonumber
  \rho_{\GAP{\rho}} &= \int\limits_{\sphere(\Hilbert)} \!\!
  \GAP{\rho} (d\psi) \, |\psi\rangle \langle\psi|= \EEE \biggl(
  |\Psi^{GAP}\rangle \langle\Psi^{GAP}| \biggr) = \\\nonumber
  &\stackrel{\eqref{mudef}}{=} \EEE \biggl( \|\Psi^{GA}\|^{-2} \,
  |\Psi^{GA}\rangle \langle\Psi^{GA}| \biggr) = \int\limits_\Hilbert
  \GA{\rho} (d\psi) \, \|\psi\|^{-2} \, |\psi \rangle \langle \psi | =
  \\\nonumber &\stackrel{\eqref{muNdef}}{=} \int\limits_\Hilbert
  \G{\rho} (d\psi) \, |\psi\rangle \langle\psi| = \rho
\end{align}
because $\int_\Hilbert \G{\rho} (d\psi) \, |\psi\rangle \langle \psi|$
is the covariance matrix of $\G{\rho}$, which is $\rho$. (A number
above an equal sign refers to the equation used to obtain the
equality.)
\end{proof}

\noindent {\bf Remark on the covariance matrix.}
The equation $\rho_{\GAP{\rho}} = \rho$ can be understood as
expressing that $\GAP{\rho}$ and $\G{\rho}$ have the same covariance.
For a probability measure $\mu$ on $\Hilbert$ with mean 0 that need
not be concentrated on $\sphere(\Hilbert)$, the covariance matrix
$C_\mu$ is given by
\begin{equation}\label{Cdef}
  C_\mu = \int\limits_\Hilbert \mu(d\phi) \, |\phi \rangle
  \langle \phi|.
\end{equation}
Suppose we want to obtain from $\mu$ a probability measure on
$\sphere(\Hilbert)$ having the same covariance. The projection
$\proj_*\mu$ of $\mu$ to $\sphere(\Hilbert)$, defined by
$\proj_*\mu(B) = \mu (\RRR^+ B)$ for $B \subseteq
\sphere(\Hilbert)$, is not what we want, as it has covariance
\[
  C_{\proj_* \mu} = \int\limits_{\sphere(\Hilbert)} \!\!
  \proj_*\mu(d\psi) \, |\psi \rangle \langle \psi| =
  \int\limits_\Hilbert \mu(d\phi) \, \|\phi\|^{-2} \, |\phi
  \rangle \langle \phi| \neq C_\mu.
\]
However, $\proj_* \bigl( A(\mu) \bigr)$ does the job: it has the same
covariance.  As a consequence, a naturally distinguished measure on
$\sphere(\Hilbert)$ with given covariance is the Gaussian adjusted
projected measure, the GAP measure, with the given covariance.

\subsection{$\GAP{\rho}$ is Covariant} 

We establish Property~2 and then discuss in
more general terms under which conditions a measure on
$\sphere(\Hilbert)$ is stationary.

\begin{proof}{ of Property~2}
  Under a unitary transformation $U$, a Gaussian measure with
  covariance matrix $C$ transforms into one with covariance matrix
  $UCU^*$. Since $\|U\psi\|^2 = \|\psi\|^2$, $\GA{C}$ transforms into
  $\GA{UCU^*}$; that is, $U\Psi^{GA}_C$ and $\Psi^{GA}_{UCU^*}$ are
  equal in distribution, and since $\|U\Psi^{GA}_C\| =
  \|\Psi^{GA}_C\|$, we have that $U\Psi_C^{GAP}$ and
  $\Psi_{UCU^*}^{GAP}$ are equal in distribution. In other words,
  $\GAP{C}$ transforms into $\GAP{UCU^*}$, which is what we claimed in
  \eqref{covariance}.
\end{proof}

\subsubsection{Stationarity} \label{sec:stationarity}
In  this subsection we discuss a criterion for stationarity
under the evolution generated by $H= \sum_n E_n \, |n \rangle \langle
n|$.  Consider the following property of a sequence of complex random
variables $Z_n$:
\begin{equation}\label{independ}
\begin{split}
  &\text{The phases $Z_n/|Z_n|$, when they exist, are independent of
    the moduli } |Z_n|\\
  &\text{and of each other, and are uniformly distributed on }S^1 =\{
    e^{i \theta}: \theta \in \RRR \}.
\end{split}
\end{equation}
(The phase $Z_n/|Z_n|$ exists when $Z_n \neq 0$.)  Condition
\eqref{independ} implies that the distribution of the random vector
$\Psi = \sum_n Z_n |n\rangle$ is stationary, since $Z_n(t) =
\exp(-iE_n t/\hbar) Z_n(0)$. Note also that \eqref{independ} implies
that the distribution has mean 0.

We show that the $Z_n = \sp{n}{\Psi^{GAP}}$ have property
\eqref{independ}. To begin with, the $Z_n = \sp{n}{\Psi^G}$ obviously
have this property since they are independent Gaussian variables.
Since the density of $\GA{\rho}$ relative to $\G{\rho}$ is a function
of the moduli alone, also the $Z_n = \sp{n}{\Psi^{GA}}$ satisfy
\eqref{independ}.  Finally, since the $|\sp{n}{\Psi^{GAP}}|$ are
functions of the $|\sp{n}{\Psi^{GA}}|$ while the phases of the
$\sp{n}{\Psi^{GAP}}$ equal the phases of the $\sp{n}{\Psi^{GA}}$, also
the $Z_n = \sp{n}{\Psi^{GAP}}$ satisfy \eqref{independ}.

We would like to add that \eqref{independ} is not merely a sufficient,
but also almost a necessary condition (and \emph{morally} a necessary
condition) for stationarity. Since for any $\Psi$, the moduli $|Z_n| =
|\sp{n}{\Psi}|$ are constants of the motion, the evolution of $\Psi$
takes place in the (possibly infinite-dimensional) torus
\begin{equation}\label{torus}
  \Big\{ \sum\limits_{n} |Z_n| e^{i \theta_n} \, |n\rangle : 0
  \le \theta_n < 2\pi \Big\} \cong \prod\limits_{n: Z_n \neq 0} S^1,
\end{equation}
contained in $\sphere(\Hilbert)$. Independent uniform phases
correspond to the uniform measure $\lambda$ on $\prod_n S^1$.
$\lambda$ is the only stationary measure if the motion on $\prod_n
S^1$ is uniquely ergodic, and this is the case whenever the spectrum
$\{ E_n\}$ of $H$ is linearly independent over the rationals $\QQQ$,
i.e., when every finite linear combination $\sum_n r_n E_n$ of
eigenvalues with rational coefficients $r_n$, not all of which vanish,
is nonzero, see \cite{ergodic,vN29}.

This is true of generic Hamiltonians, so that $\lambda$ is generically
the unique stationary distribution on the torus. But even when the
spectrum of $H$ is linearly dependent, e.g.\ when there are degenerate
eigenvalues, and thus further stationary measures on the torus exist,
these further measures should not be relevant to thermal equilibrium
measures, because of their instability against perturbations of $H$
\cite{HKTP74,LAG75}.

The stationary measure $\lambda$ on $\prod_n S^1$ corresponds, for
given moduli $|Z_n|$ or, equivalently, by setting $|Z_n| =
p(E_n)^{1/2}$ for a given probability measure $p$ on the spectrum of
$H$, to a stationary measure $\lambda_p$ on $\sphere(\Hilbert)$ that
is concentrated on the embedded torus \eqref{torus}. The measures
$\lambda_p$ are (for generic $H$) the extremal stationary measures,
i.e., the extremal elements of the convex set of stationary measures,
of which all other stationary measures are mixtures.

\subsection{GAP Measures and Gaussian Measures}\label{gmgm}

Lemma~\ref{GaussGAP} is more or less immediate from the definition of
$\GAP{\rho}$. A more detailed proof looks like this:

\begin{proof}{ of Lemma~\ref{GaussGAP}}
  By assumption the distribution $\mu \circ \Psi^{-1}$ of $\Psi$ with
  respect to $\mu$ is $\G{\rho}$. Thus for the distribution of $\Psi$
  with respect to $\mu' (d\omega) = \|\Psi(\omega)\|^2 \mu (d\omega)$,
  we have $\mu' \circ \Psi^{-1} (d\psi) = \|\psi\|^2 \, \mu \circ
  \Psi^{-1} (d\psi) = \|\psi\|^2 \, \G{\rho} (d\psi) = \GA{\rho}
  (d\psi)$.  Thus, $\proj(\Psi(\omega))$ has distribution $\proj_*
  \GA{\rho} = \GAP{\rho}$.
\end{proof}

\subsection{Generalized Bases}

We have already remarked in the introduction that the
orthonormal basis $\{|q_2\rangle\}$ of $\Hilbert_2$, used in the
definition of the conditional wave function, could be a
\emph{generalized} basis, such as a ``continuous'' basis, for which it
is appropriate to write
\[
  I_2 = \int dq_2 \, |q_2 \rangle \langle q_2|
\]
instead of the ``discrete'' notation
\[
 I_2  = \sum_{q_2} |q_2 \rangle \langle q_2|
\]
we used in \eqref{Psi1def}--\eqref{rhoreddef}. 

We wish to elucidate this further. A generalized orthonormal basis
$\{|q_2 \rangle: q_2 \in \conf_2\}$ indexed by the set $\conf_2$ is
mathematically defined by a unitary isomorphism $\Hilbert_2 \to
L^2(\conf_2,dq_2)$, where $dq_2$ denotes a measure on $\conf_2$. We
can think of $\conf_2$ as the configuration space of system 2; as a
typical example, system 2 may consist of $N_2$ particles in a box
$\Lambda \subset \RRR^3$, so that its configuration space is $\conf_2
= \Lambda^{N_2}$ with $dq_2$ the Lebesgue measure (which can be
regarded as obtained by combining $N_2$ copies of the volume measure
on $\RRR^3$).\footnote{In fact, in the original definition of the
  conditional wave function in \cite{DGZ}, $q_2$ was supposed to be
  the \emph{configuration}, corresponding to the positions of the
  particles belonging to system 2.  For our purposes here, however,
  the physical meaning of the $q_{2}$ is irrelevant, so that any
  generalized orthonormal basis of $\Hilbert_{2}$ can be used.} The
formal ket $|q_2 \rangle$ then means the delta function centered at
$q_2$; it is to be treated as (though strictly speaking it is not) an
element of $\Hilbert_2$.

The definition of the conditional wave function $\Psi_1$ then reads as
follows: The vector $\psi \in \Hilbert_1 \otimes \Hilbert_2$ can be
regarded, using the isomorphism $\Hilbert_2 \to L^2(\conf_2,dq_2)$, as
a function $\psi : \conf_2 \to \Hilbert_1$.  Eq.~\eqref{Psi1def} is
to be understood as meaning
\begin{equation}\label{Psi1defcont}
  \Psi_1 = \mathcal{N} \: \psi(Q_2)
\end{equation}
where 
\[
  \mathcal{N}= \mathcal{N} (\psi,Q_2) = \bigl\|\psi(Q_2) \bigr\|^{-1}
\]
is the  normalizing factor and $Q_2$ is a random point in $\conf_2$,
chosen with the quantum distribution
\begin{equation}\label{margcont}
  \PPP(Q_2 \in dq_2) = \bigl\| \psi(q_2) \bigr\|^2 dq_2 \,,
\end{equation}
which is how \eqref{marg} is to be understood in this setting.  As
$\psi$ is defined only up to changes on a null set in $\conf_2$,
$\Psi_1$ may not be defined for a particular $Q_2$. Its distribution
in $\Hilbert_1$, however, is defined unambiguously by
\eqref{Psi1defcont}. In the most familiar setting with $\Hilbert_1 =
L^2(\conf_1,dq_1)$, we have that $(\psi(Q_2))(q_1) = \psi(q_1,Q_2)$.

In the following, we will allow generalized bases and use continuous
instead of discrete notation, and set
$\sp{Q_2}{\psi} = \psi(Q_2)$.

\subsection{Distribution of the Wave Function of a Subsystem} 
\label{sec:sub}

\begin{proof}{ of Property~3}
The proof is divided into four steps.

\textit{Step 1.}  We can assume that $\Psi = \proj (\Psi^{GA})$ where
$\Psi^{GA}$ is a $\GA{\rho}$-distributed random vector in $\Hilbert =
\Hilbert_1 \otimes \Hilbert_2$. We then have that $\Psi_1 = \proj_1
\bigl( \sp{Q_2}{\Psi} \bigr) = \proj_1 \bigl( \sp{Q_2}{ \Psi^{GA}}
\bigr)$ where $\proj_1$ is the normalization in $\Hilbert_1$, and
where the distribution of $Q_2$, given $\Psi^{GA}$, is
\[
  \PPP(Q_2 \in dq_2|\Psi^{GA}) = \frac{\| \sp{q_2}{\Psi^{GA}} \|^2}{\|
  \Psi^{GA} \|^2} dq_2\,.
\]
$\Psi^{GA}$ and $Q_2$ have a joint distribution given
by the following measure $\nu$ on $\Hilbert \times \conf_2$:
\begin{equation}
  \nu(d\psi \times dq_2) = \| \sp{q_2}{\psi} \|^2 \, \G{\rho}(d\psi)
  \, dq_2 \,.
\end{equation}
Thus, what needs to be shown is that with respect to $\nu$, $\proj_1
(\sp{q_2}{\psi})$ is $\GAP{\rho_1}$-distributed.

\textit{Step 2.} \emph{If $\Psi \in \Hilbert_1 \otimes \Hilbert_2$ is
  $\G{\rho_1 \otimes \rho_2}$-distributed and $q_2 \in \conf_2$ is
  fixed, then the random vector $f(q_2) \, \sp{q_2}{\Psi} \in
  \Hilbert_1$ with $f(q_2) = \langle q_2 | \rho_2 | q_2
  \rangle^{-1/2}$ is $\G{\rho_1}$-distributed.}  This follows, more or
less, from the fact that a subset of a set of jointly Gaussian random
variables is also jointly Gaussian, together with the observation that
the covariance of $\sp{q_2}{\Psi}$ is 
\[
\int_\Hilbert \G{\rho_1 \otimes \rho_1} (d\psi) \,
\sp{q_2}{\psi}\sp{\psi}{q_2} = \sp{q_2} {\rho_1 \otimes \rho_2| q_2} =
\rho_1 \, \sp{q_2}{\rho_2|q_2}\,.
\]
More explicitly, pick an orthonormal basis $\{|n_i \rangle \}$ of
$\Hilbert_i$ consisting of eigenvectors of $\rho_i$ with eigenvalues
$p^{(i)}_{n_i}$, and note that the vectors $| n_1,n_2 \rangle:= |n_1
\rangle \otimes |n_2 \rangle$ form an orthonormal basis of $\Hilbert =
\Hilbert_1 \otimes \Hilbert_2$ consisting of eigenvectors of $\rho_1
\otimes \rho_2$ with eigenvalues $p_{n_1,n_2} = p^{(1)}_{n_1}
p^{(2)}_{n_2}$. Since the random variables $Z_{n_1,n_2} := \sp{
  n_1,n_2 }{ \Psi}$ are independent Gaussian random variables with
mean zero and variances $\EEE|Z_{n_1},{n_2}|^2 = p_{n_1,n_2}$, so are
their linear combinations
\[
  Z_{(1)n_1} := \sp{ n_1 }{ f(q_2) \, \Psi(q_2) } = f(q_2) \sum_{n_2}
  \sp{ q_2 }{ n_2} \, Z_{n_1,n_2}
\]
with variances (because variances add when adding independent Gaussian
random variables)
\[
  \EEE |Z_{(1)n_1}|^2 = f^2(q_2) \sum_{n_2} \bigl| \sp{q_2}{ n_2}
  \bigr|^2 \EEE |Z_{n_1,n_2}|^2 =
  p^{(1)}_{n_1} \frac{ \sum_{n_2} |\sp{q_2}{n_2}|^2 \, p^{(2)}_{n_2}}
  {\sp{q_2}{ \rho_2 | q_2}} = p^{(1)}_{n_1}.
\]
Thus $f(q_2) \, \sp{q_2}{\Psi}$ is $\G{\rho_1}$-distributed, which
completes step 2.

\textit{Step 3.} \emph{If $\Psi \in \Hilbert_1 \otimes \Hilbert_2$ is
  $\G{\rho_1 \otimes \rho_2}$-distributed and $Q_2 \in \conf_2$ is
  random with any distribution, then the random vector $f(Q_2) \,
  \sp{Q_2}{\Psi}$ is $\G{\rho_1}$-distributed.} This is a trivial
consequence of step 2.

\textit{Step 4.} Apply Lemma~\ref{GaussGAP} as follows. Let $\Omega =
\Hilbert \times \conf_2$, $\Psi(\omega) = \Psi(\psi, q_2) = f(q_2) \,
\sp{q_2}{\psi}$, and $\mu(d\psi \times dq_2) = \G{\rho}(d\psi) \,
\sp{q_2}{\rho_2|q_2} \, dq_2$ (which means that $q_2$ and $\psi$ are
independent). By step 3, the hypothesis of Lemma~\ref{GaussGAP} (for
$\rho = \rho_1$) is satisfied, and thus $\proj_1 (\Psi) = \proj_1
(\sp{q_2}{\psi})$ is $\GAP{\rho_1}$-distributed with respect to
\[
  \|\Psi(\omega)\|^2 \mu(d\omega) = 
  {f^2(q_2)} \|\sp{q_2}{\psi}\|^2 \, \G{\rho}(d\psi) \,
  \sp{q_2}{\rho_2|q_2} \, dq_2 = \nu(d\omega)\,,
\] 
where we have used that $f^2(q_2)=\langle q_2|\rho_2|q_2 \rangle^{-1}$.
But this is, according to step 1, what we needed to show. 
\end{proof}


To verify the statement after Property~3, consider the density
matrix $\rho = |\Phi \rangle \langle \Phi|$ for a pure state $\Phi$ of
the form $\Phi = \sum_n \sqrt{p_n} \, \psi_n \otimes \phi_n$, where
$\{\psi_n\}$ and $\{\phi_n\}$ are respectively orthonormal bases for
$\Hilbert_1$ and $\Hilbert_2$ and the $p_n$ are nonnegative with
$\sum_n p_n =1$. Then a $\GAP{\rho}$-distributed random vector $\Psi$
coincides with $\Phi$ up to a random phase, and so $\rho_1^\red =
\sum_n p_n \, |\psi_n \rangle \langle \psi_n|$.  Choosing for $\{|q_2
\rangle\}$ the basis $\{\phi_n\}$, the distribution of $\Psi_1$ is not
$\GAP{\rho_1^\red}$ but rather is concentrated on the eigenvectors of
$\rho_1^\red$. When the $p_n$ are pairwise-distinct this measure is the
measure $\EIG{\rho_1^\red}$ we define in Section~\ref{sec:nu}.


\section{Microcanonical Distribution for a Large System 
  Implies the  Distribution $\GAP{\rho_\beta}$ for a Subsystem}
\label{sec:hb1}

In this section we use Property~3, i.e., the fact that GAP measures are
hereditary, to show that $\GAP{\rho_\beta}$ is the distribution of the
conditional wave function of a system coupled to a heat bath when the wave
function of the composite is distributed microcanonically, i.e., according
to $u_{E,\delta}$.

Consider a system with Hilbert space $\Hilbert_1$ coupled to a heat
bath with Hilbert space $\Hilbert_2$. Suppose the composite system has
a random wave function $\Psi \in \Hilbert = \Hilbert_1 \otimes
\Hilbert_2$ whose distribution is microcanonical, $u_{E,\delta}$.
Assume further that the coupling is negligibly small, so that we can
write for the Hamiltonian
\begin{equation}\label{nocoupling}
  H = H_1 \otimes I_2 + I_1 \otimes H_2 \,,
\end{equation}
and that the heat bath is large (so that the energy levels of
$H_2$ are very close).  


\z{It is a well known fact that for macroscopic systems different
equilibrium ensembles, for example the microcanonical and the canonical, give
approximately the same answer for appropriate quantities.} By \z{this}
equivalence of ensembles \cite{martinlof}, we should have that
$\rho_{E,\delta} \approx \rho_\beta$ for suitable $\beta = \beta(E)$.
Then, since $\GAP{\rho}$ depends continuously on $\rho$, we have that
$u_{E,\delta} = \GAP{\rho_{E,\delta}} \approx \GAP{\rho_\beta}$. Thus we
should have that the distribution of the conditional wave function $\Psi_1$
of the system is approximately the same as would be obtained when $\Psi$ is
$\GAP{\rho_\beta}$-distributed. But since, by \eqref{nocoupling}, the
canonical density matrix is then of the form
\begin{equation}
  \rho_\beta = \rho_{\Hilbert,H,\beta} = \rho_{\Hilbert_1, H_1, \beta} \otimes
  \rho_{\Hilbert_2, H_2, \beta}\,,
\end{equation}
we have by Property~3 that $\Psi_1$ is approximately 
$\GAP{\rho_{\Hilbert_1, H_1, \beta}}$-distributed, which is what we
wanted to show.

\section{Typicality of GAP Measures}
\label{sec:Typicality}

The previous section concerns the distribution of the conditional wave
function $\Psi_1$ arising from the microcanonical distribution of the wave
function of the composite. It concerns, in other words, a \emph{random}
wave function of the composite. The result there is the analogue, on the
level of measures on Hilbert space, of the well known result that if a
microcanonical density matrix
\eqref{microcanrho} is assumed for the composite, the reduced density
matrix $\rho_1^\red$ of the system, defined as the partial trace $\tr_2 \,
\rho_{E,\delta}$, is canonical if the heat bath is large \cite{LL59}.  

As indicated in the introduction, a stronger statement about the canonical
density matrix is in fact true, namely that for a
\emph{fixed} (nonrandom) wave function $\psi$ of the composite which
is typical with respect to $u_{E,\delta}$, $\rho_1^\red \approx
\rho_{\Hilbert_1, H_1, \beta}$ when the heat bath is large (see
\cite{thermo4, schrbook}; for a rigorous study of special cases of
a similar question, see \cite{Tasaki1}).\footnote{It is a consequence of
  the results in \cite{Pag93} that when $\dim\Hilbert_2\to\infty$, the
  reduced density matrix becomes proportional to the identity on
  $\Hilbert_1$ for typical wave functions relative to the uniform
  distribution on $\sphere(\Hilbert)$ (corresponding to $u_{E,\delta}$ for
  $E=0$ and  $H=0$).}
This stronger statement will be used in Section~\ref{sec:hb2} to show that
a similar statement holds for the distribution of $\Psi_1$ as well, namely
that it is approximately $\GAP{\rho_{\Hilbert_1, H_1, \beta}}$-distributed
for a typical fixed $\psi
\in \Hilbert_{E,\delta}$ and basis \onb\ of $\Hilbert_2$. But we must first
consider the distribution of $\Psi_1$ for a typical $\psi\in \Hilbert$.

\subsection{Typicality of GAP Measures for a Subsystem
  of a Large System}
\label{sec:typicality}

In this section we argue that for a typical wave function of a big system
the conditional wave function of a small subsystem is approximately
GAP-distributed, first giving a precise formulation of this result and then
sketching its proof. We give the detailed proof in \cite{thermo6}.

\subsubsection{Statement of the Result}

Let $\Hilbert = \Hilbert_1 \otimes \Hilbert_2$, where
$\Hilbert_1$ and $\Hilbert_2$ have respective 
dimensions $k$ and $m$, with $k < m < \infty$.  For any given density matrix
$\rho_1$ on $\Hilbert_1$, consider the  set
\begin{equation}\label{Rdef}
  \R (\rho_1) = \bigl\{\psi \in \sphere(\Hilbert): \rho_1^\red (\psi) =
  \rho_1 \bigr\} \,,
\end{equation}
where $\rho_1^\red(\psi) = \tr_2 |\psi \rangle \langle \psi|$ is the
reduced density matrix for the wave function $\psi$.  There is a natural notion of (normalized)
uniform measure $u_{\rho_1}$ on $\R(\rho_1)$; we give
its precise definition in Section~\ref{sec:typoutline}.

We claim that for fixed $k$ and large $m$, the distribution $\mu_1^{\psi}$
of the conditional wave function $\Psi_1$ of system 1, defined by
\eqref{Psi1def} and \eqref{marg} for a basis $\{ |q_2\rangle\}$ of
$\Hilbert_2$, is close to $\GAP{\rho_1}$ for the
overwhelming majority, relative to $u_{\rho_1}$, of vectors $\psi \in \Hilbert$
with reduced density matrix $\rho_1$. More precisely:

\bigskip

\noindent\textit{For every   $\varepsilon > 0$ and every bounded continuous function $f:
  \sphere(\Hilbert_1) \to \RRR$,
  \begin{equation}\label{GAPtyp}
    u_{\rho_1} \Bigl\{\psi \in \R(\rho_1): \bigl|
    \mu_1^{\psi}(f) - \GAP{\rho_1}(f) \bigr| < \varepsilon \Bigr\} \to
    1 \quad \text{as } \, m \to \infty \,,
  \end{equation}
regardless of how the basis $\{ |q_2\rangle\}$ is chosen.} 

\bigskip

\noindent Here we use the notation
\begin{equation}
  \mu(f) := \int\limits_{\sphere(\Hilbert)} \mu(d\psi)\,
  f(\psi) \,.
\end{equation}



\subsubsection{Measure on $\Hilbert$ Versus Density Matrix}

It is important to resist the temptation to translate $u_{\rho_1}$ into a
density matrix in $\Hilbert$.  As mentioned in the introduction, to every
probability measure $\mu$ on $\sphere(\Hilbert)$ there corresponds a
density matrix $\rho_\mu$ in $\Hilbert$, given by
\eqref{rhomupsi}, which contains all the empirically accessible information
about an ensemble with distribution $\mu$. It may therefore seem a natural
step to consider, instead of the measure $\mu = u_{\rho_1}$, directly its
density matrix $\rho_\mu=\frac1m\rho_1\otimes I_2$, where $I_2$ is the
identity on $\Hilbert_2$. But since our result concerns properties of most
wave functions relative to $\mu$, it cannot be formulated in terms of the
density matrix $\rho_\mu$. In particular, the corresponding statement
relative to another measure $\mu' \neq \mu$ on $\sphere(\Hilbert)$ with the
same density matrix $\rho_{\mu'} = \rho_\mu$ could be false. Noting that
$\rho_\mu$ has a basis of eigenstates that are product vectors, we could,
for example, take $\mu'$ to be a measure concentrated on these eigenstates.
For any such state $\psi$, $\mu_1^{\psi}$ is a delta-measure.


\subsubsection{Outline of Proof}
\label{sec:typoutline}

The result follows, by \eqref{marg}, Lemma~\ref{GaussGAP}, and the
continuity of $\proj_* \circ A$, from the corresponding statement about the
Gaussian measure $\G{\rho_1}$ on $\Hilbert_1$ with covariance $\rho_1$:

\bigskip

\noindent\textit{For every $\varepsilon > 0$ and every
  bounded continuous $f: \Hilbert_1 \to \RRR$,}
\begin{equation}\label{Gtyp}
  u_{\rho_1} \Bigl\{\psi \in \R(\rho_1): \bigl| \bar
  \mu_1^{\psi}(f) - \G{\rho_1}(f) \bigr| < \varepsilon \Bigr\} \to 1
  \quad \text{\textit{as} } \, m \to \infty \,,
\end{equation}
\textit{where $\bar \mu_1^\psi$ is the distribution of $\sqrt{m}
  \, \sp{Q_2}{\psi} \in \Hilbert_1$ (not normalized) with respect to
  the uniform distribution of $Q_2 \in \{1, \ldots, m\}$.}

\bigskip

We sketch the proof of \eqref{Gtyp} and give the definition of $u_{\rho_1}$.
According to the Schmidt decomposition, every $\psi \in \Hilbert$ can
be written in the form
\begin{equation}\label{Schmidt}
  \psi = \sum_i c_i \, \chi_i \otimes \phi_{i} \,,
\end{equation}
where $\{\chi_i \}$ is an orthonormal basis of 
$\Hilbert_{1}$, $\{\phi_i\}$ an orthonormal system in $\Hilbert_2$,
and the $c_i$ are coefficients which can be assumed real and
nonnegative. {}From \eqref{Schmidt} one reads off the reduced density
matrix of system 1,
\begin{equation}\label{Schmidtrho}
  \rho_1^\red = \sum_{i} c_i^2 \, |\chi_{i} \rangle \langle
  \chi_{i} |\,.
\end{equation}
As the reduced density matrix is given, $\rho_1^\red = \rho_1$, the
orthonormal basis $\{\chi_{i} \}$ and the coefficients $c_i$ are
determined (when $\rho_1$ is nondegenerate) as the eigenvectors and the
square-roots of the eigenvalues of $\rho_1$, and, $\R(\rho_1)$ is in a
natural one-to-one correspondence with the set $ONS(\Hilbert_2,k)$ of all
orthonormal systems $\{\phi_{i} \}$ in $\Hilbert_2$ of cardinality $k$. (If
some of the eigenvalues of $\rho_1$ vanish, the one-to-one correspondence
is with $ONS(\Hilbert_2,k')$ where $k'=\dim\support(\rho_1)$.)  The Haar
measure on the unitary group of $\Hilbert_2$ defines the uniform
distribution on the set of orthonormal bases of $\Hilbert_2$, of which the
uniform distribution on $ONS(\Hilbert_2,k)$ is a marginal, and thus defines
the uniform distribution $u_{\rho_1}$ on $\R(\rho_1)$. (When $\rho_1$ is
degenerate, $u_{\rho_1}$ does not depend upon how the eigenvectors $\chi_{i}$ of
$\rho_1$ are chosen.)

The key idea for establishing \eqref{Gtyp} from the Schmidt decomposition
\eqref{Schmidt} is this: $\bar\mu_1^\psi$ is the average of $m$ delta
measures with equal weights, $\bar\mu_1^\psi = m^{-1} \sum_{q_2}
\delta_{\psi_1(q_2)}$, located at the points
\begin{equation}
  \psi_1(q_2) = \sum_{i=1}^{k} c_i \, \sqrt{m} \, \sp{q_2}{\phi_{i}}
  \, \chi_{i} \,.
\end{equation}
Now regard $\psi$ as random with distribution $u_{\rho_1}$; then
the $\psi_1(q_2)$ are $m$ random vectors, and $\bar\mu^\psi_1$ is
their empirical distribution.  If the $mk$ coefficients
$\sp{q_2}{\phi_{i}}$ were \emph{independent Gaussian} (complex) random
variables with (mean zero and) variance $m^{-1}$, then the
$\psi_1(q_2)$ would be $m$ independent drawings of a
$\G{\rho_1}$-distributed random vector; by the weak law of large
numbers, their empirical distribution would usually be close to $\G{\rho_1}$;
in fact, the probability that $\bigl| \bar\mu_1^{\psi}(f) -
\G{\rho_1}(f) \bigr| < \varepsilon$ would converge to 1, as $m \to
\infty$.

However, when $\{\phi_{i}\}$ is a random orthonormal system with uniform
distribution as described above, the expansion coefficients
$\sp{q_2}{\phi_{i}}$ in the decomposition of the $\phi_{i}$'s
\begin{equation}\label{expand}
\phi_i=\sum_{q_2} \sp{q_2}{\phi_{i}}|q_2\rangle 
\end{equation}
will not be independent---since the $\phi_{i}$'s must be orthogonal and
since $\|\phi_i\|=1$. Nonetheless, replacing the coefficients
$\sp{q_2}{\phi_{i}}$ in
\eqref{expand} by independent Gaussian coefficients $a_i(q_2)$ as described
above, we obtain a system of vectors 
\begin{equation}
\phi_i'=\sum_{q_2} a_i(q_2)|q_2\rangle 
\end{equation}
that, in the limit $m\to\infty$, form a uniformly distributed orthonormal
system: $\|\phi_i'\|\to1$ (by the law of large numbers) and
$\sp{\phi_i'}{\phi_j'}\to0$ for $i\neq j$ (since a pair of randomly chosen
vectors in a high-dimensional Hilbert space will typically be almost
orthogonal). This completes the proof.

\subsubsection{Reformulation}\label{sec:ref}
 
While this result {\it suggests} that $\GAP{\rho_\beta}$ is the
distribution of the conditional wave function of a system coupled to a heat
bath when the wave function of the composite is a typical {\it fixed}
microcanonical wave function, belonging to $\Hilbert_{E,\delta}$, it does
not quite {\it imply} it. The reason for this is that $\Hilbert_{E,\delta}$
has measure 0 with respect to the uniform distribution on $\Hilbert,$ even
when the latter is finite-dimensional. Nonetheless, there is a simple
corollary, or reformulation, of the result that will allow us to cope with
microcanonical wave functions.

We have indicated that for our result the choice of basis \onb\ of
$\Hilbert_2$ does not matter. In fact, while  $\mu_1^{\psi}$, the
distribution of the conditional wave function  $\Psi_1$ of system 1,
depends upon both $\psi\in\Hilbert$ and the choice of basis \onb\ of
$\Hilbert_2$, the distribution of $\mu_1^{\psi}$ itself, when $\psi$ is
$u_{\rho_1}$-distributed, does not depend upon the choice of basis. This follows
from the fact that for any unitary $U$ on $\Hilbert_2$
\begin{equation}\label{U}
\sp{U^{-1}q_2}{\psi}=\sp{q_2}{I_1\otimes U\psi}
\end{equation}
(and the invariance of the Haar measure of the unitary group of
$\Hilbert_2$ under left multiplication). It similarly follows from
\eqref{U} that for fixed $\psi\in\Hilbert$, the distribution of
$\mu_1^{\psi}$ arising from the uniform distribution $\nu$ of the basis
\onb, in the set $ONB(\Hilbert_2)$ of all orthonormal bases of
$\Hilbert_2$, is the same as the distribution of  $\mu_1^{\psi}$ 
arising from the uniform distribution $u_{\rho_1}$ of $\psi$ with a fixed
basis (and the fact that the Haar measure is invariant under $U\mapsto
U^{-1}$). We thus have the following corollary:

\bigskip

\noindent
\textit{Let $\psi\in\Hilbert$ and let $\rho_1=\tr_2 |\psi
  \rangle \langle \psi|$ be the corresponding reduced density matrix for
  system 1. Then for a typical basis \onb\ of $\Hilbert_2$, the conditional
  wave function  $\Psi_1$ of system 1 is approximately
  $\GAP{\rho_1}$-distributed when $m$ is large: 
For every   $\varepsilon > 0$ and every bounded continuous function $f:
  \sphere(\Hilbert_1) \to \RRR$,
  \begin{equation}\label{GAPtyp2}
    \nu \Bigl\{\{ |q_2\rangle\}\in ONB(\Hilbert_2): \bigl|
    \mu_1^{\psi}(f) - \GAP{\rho_1}(f) \bigr| < \varepsilon \Bigr\} \to
    1 \quad \text{as } \, \dim(\Hilbert_2) \to \infty \,.
  \end{equation}
}

\subsection{Typicality of $\GAP{\rho_\beta}$ for a
  Subsystem of a Large System in the Microcanonical Ensemble}
\label{sec:hb2}


It is an immediate consequence of the result of Section~\ref{sec:ref} that
for any fixed microcanonical wave function $\psi$ for a system coupled to a
(large) heat bath, the conditional wave function $\Psi_1$ of the system will be
approximately GAP-distributed. When this is combined with the ``canonical
typicality'' described near the beginning of Section~\ref{sec:Typicality}, we
obtain the following result:

\bigskip

\noindent
\emph{Consider a system with finite-dimensional Hilbert space $\Hilbert_1$
  coupled to a   heat bath with finite-dimensional Hilbert space
  $\Hilbert_2$. Suppose   that the coupling 
  is weak, so that we can write $H = H_1 \otimes I_2+ I_1\otimes H_2$ on
  $\Hilbert=\Hilbert_1\otimes\Hilbert_2$,  and
  that the heat bath is large, so that the eigenvalues of $H_2$ are
  close. Then for any wave function $\psi$ that is typical relative to
  the microcanonical measure 
  $u_{E,\delta}$, the distribution $\mu_1^{\psi}$ of the conditional
  wave function $\Psi_1$, defined by \eqref{Psi1def} and \eqref{marg} for a
  typical basis $\{ |q_2\rangle\}$ of the heat bath,
  is close to $\GAP{\rho_\beta}$ for suitable $\beta=\beta(E)$, where
  $\rho_\beta=\rho_{\Hilbert_1, H_1, \beta}$. In other words,
  in the thermodynamic limit, in which the volume $V$ of the heat bath and
  $dim(\Hilbert_2)$ go to infinity and  $E/V=e$ is constant, we have that  for
  all $\varepsilon, \delta >0$, 
  and for all bounded continuous functions $f: \sphere(\Hilbert_1) \to
  \RRR$,
  \begin{equation}\label{microcan}
    u_{E,\delta}\times\nu \Bigl\{(\psi,\{|q_2\rangle\}) \in
    \sphere(\Hilbert)\times ONB(\Hilbert_2) : \bigl| 
    \mu_1^{\psi}(f) - \GAP{\rho_\beta}(f) \bigr| < \varepsilon \Bigr\} \to
    1
  \end{equation}
where $\beta=\beta(e)$.
}

We note that if $\{ |q_2\rangle\}$ were an energy eigenbasis rather
than a typical basis, the result would be false.


\section{Remarks}\label{sec:rem}

\subsection{Other Candidates for the Canonical Distribution}

We review in this section other distributions that have been, or may be,
considered as possible candidates for the distribution of the wave function
of a system from a canonical ensemble.

\subsubsection{A Distribution on the Eigenvectors}\label{sec:nu}

One possibility, which goes back to von Neumann
\cite[p.~329]{Neumann}, is to consider $\mu(d\psi)$ as concentrated on
the eigenvectors of $\rho$; we denote this distribution $\EIG{\rho}$
after the first letters of ``eigenvector''; it is defined as follows.
Suppose first that $\rho$ is nondegenerate. To select an
$\EIG{\rho}$-distributed vector, pick a unit eigenvector $|n \rangle$,
so that $\rho|n\rangle = p_n |n\rangle$, with probability $p_n$ and
randomize its phase. This definition can be extended in a natural way
to degenerate $\rho$:
\begin{equation}\label{nudef}
  \EIG{\rho} = \sum_{p \in \spec (\rho)} p \, \dim \Hilbert_p \:\:
  u_{\sphere(\Hilbert_p)},
\end{equation}
where $\Hilbert_p$ denotes the eigenspace of $\rho$ associated with
eigenvalue $p$.  The measure $\EIG{\rho}$ is concentrated on the set
$\bigcup_p \Hilbert_p$ of eigenvectors of $\rho$, which for the
canonical $\rho = \rho_{\Hilbert, H, \beta}$ coincides with the set of
eigenvectors of $H$; it is a mixture of the microcanonical
distributions $u_{\sphere(\Hilbert_p)}$ on the eigenspaces of $H$ in
the same way as in classical mechanics the canonical distribution on
phase space is a mixture of the microcanonical distributions.

Note that $\EIG{\rho_{E,\delta}}=u_{E,\delta}$, and that in particular
$\EIG{\rho_{E,\delta}}$ is not, when $H$ is nondegenerate, the uniform
distribution $\mu_{E,\delta}$ on the energy eigenstates with energies in
$[E,E+\delta]$, against which we have argued in the introduction. 

The distribution $\EIG{\rho}$ has the same properties as those of
$\GAP{\rho}$ described in Properties~1--3, except when
$\rho$ is degenerate:

\textit{
  The measures $\EIG{\rho}$ are such that (a) they have the right
  density matrix: $\rho_{\EIG{\rho}} = \rho$; (b) they are
  covariant: $U_* \EIG{\rho} = \EIG{U\rho U^*}$; (c) they are
  hereditary at nondegenerate $\rho$: when $\Hilbert = \Hilbert_1
  \otimes \Hilbert_2$ and $\rho$ is nondegenerate and uncorrelated,
  $\rho = \rho_1 \otimes \rho_2$, then $\EIG{\rho}$ has marginal
  (i.e., distribution of the conditional wave function)
  $\EIG{\rho_1}$.
}

\begin{proof}{}
(a) and (b) are obvious. For (c) let, for
$i=1,2$, $|n_i\rangle$ be a basis consisting of eigenvectors of
$\rho_i$ with eigenvalues $p^{(i)}_{n_i}$. Note that the tensor
products $|n_1\rangle \otimes |n_2 \rangle$ are eigenvectors of $\rho$
with eigenvalues $p^{(1)}_{n_1} p^{(2)}_{n_2}$, and by nondegeneracy
all eigenvectors of $\rho$ are of this form up to a phase factor.
Since an $\EIG{\rho}$-distributed random vector $\Psi$ is almost
surely an eigenvector of $\rho$, we have $\Psi = e^{i\Theta}
|N_1\rangle |N_2 \rangle$ with random $N_1$, $N_2$, and $\Theta$. The
conditional wave function $\Psi_1$ is, up to the phase, the
eigenvector $|N_1 \rangle$ of $\rho_1$ occurring as the first factor
in $\Psi$. The probability of obtaining $N_1 = n_1$ is $\sum_{n_2}
p^{(1)}_{n_1} p^{(2)}_{n_2} = p^{(1)}_{n_1}$.\footnote{The relevant
  condition for (c) follows from nondegeneracy but is weaker: it is
  that the eigenvalues of $\rho_1$ and $\rho_2$ are multiplicatively
  independent, in the sense that $p^{(1)}_{n_1} p^{(2)}_{n_2} =
  p^{(1)}_{m_1} p^{(2)}_{m_2}$ can occur only trivially, i.e., when
  $p^{(1)}_{n_1}=p^{(1)}_{m_1}$ and $p^{(2)}_{n_2}=p^{(2)}_{m_2}$. In
  particular, the nondegeneracy of $\rho_1$ and $\rho_2$ is
  irrelevant.}
\end{proof}

In contrast, for a \emph{degenerate} $\rho = \rho_1 \otimes \rho_2$
the conditional wave function need not be $\EIG{\rho_1}$-distributed,
as the following example shows. Suppose $\rho_1$ and $\rho_2$ are
nondegenerate but $p^{(1)}_{n_1} p^{(2)}_{n_2} = p^{(1)}_{m_1}
p^{(2)}_{m_2}$ for some $n_1\neq m_1$; then an
$\EIG{\rho}$-distributed $\Psi$, whenever it happens to be an
eigenvector associated with eigenvalue $p^{(1)}_{n_1} p^{(2)}_{n_2}$,
is of the form $c|n_1\rangle |n_2\rangle + c' |m_1 \rangle |m_2
\rangle$, almost surely with nonvanishing coefficients $c$ and $c'$;
as a consequence, the conditional wave function is a multiple of
$c|n_1\rangle \langle Q_2|n_2\rangle + c' |m_1 \rangle \langle Q_2|m_2
\rangle$, which is, for typical $Q_2$ and unless $|n_2\rangle$ and
$|m_2 \rangle$ have disjoint supports, a nontrivial superposition of
eigenvectors $|n_1\rangle$, $|m_1\rangle$ with different
eigenvalues---and thus cannot arise from the $\EIG{\rho_1}$
distribution.\footnote{A property weaker than (c) does hold for
  $\EIG{\rho}$ also in the case of the degeneracy of $\rho = \rho_1
  \otimes \rho_2$: if the orthonormal basis $\{ |q_2\rangle\}$ used in
  the definition of conditional wave function consists of eigenvectors
  of $\rho_2$, then the distribution of the conditional wave function
  is $\EIG{\rho_1}$.}

Note also that $\EIG{\rho}$ is discontinuous as a function of $\rho$ at
every degenerate $\rho$; in other words, $\EIG{\rho_{\Hilbert, H, \beta}}$
is, like $\mu_{E,\delta}$, unstable against small perturbations of the
Hamiltonian. (And, as with $\mu_{E,\delta}$, this fact, quite independently
of the considerations on behalf of GAP-measures in Sections~\ref{sec:hb1}
and \ref{sec:Typicality}, suggests against using $\EIG{\rho_\beta}$ as a
thermal equilibrium distribution.) Moreover, $\EIG{\rho}$ is highly
concentrated, generically on a one-dimensional subset of
$\sphere(\Hilbert)$, and in the case of a finite-dimensional Hilbert space
$\Hilbert$ fails to be absolutely continuous relative to the uniform
distribution $u_{\sphere (\Hilbert)}$ on the unit sphere.

For further discussion of families $\mu(\rho)$ of measures satisfying the
analogues of Properties~1--3, see
Section~\ref{sec:uniqueness}.  

\subsubsection{An Extremal Distribution}

Here is another distribution on $\Hilbert$ associated with the density
matrix $\rho$. Let the random vector $\Psi$ be
\begin{equation}\label{anotherdef}
  \Psi = \sum_{p \in \spec(\rho)} \sqrt{p} \, \Psi_p,
\end{equation}
the $\Psi_p$ being independent random vectors with distributions
$u_{\sphere(\Hilbert_p)}$. In case all eigenvalues are nondegenerate,
this means the coefficients $Z_n$ of $\Psi$, $\Psi = \sum_n Z_n
|n\rangle$, have independent uniform phases but fixed moduli $|Z_n| =
\sqrt{p_n}$---in sharp contrast with the moduli when $\Psi$ is
$\GAP{\rho}$-distributed. And in contrast to the measure $\EIG{\rho}$
considered in the previous subsection, the weights $p_n$ in the
density matrix now come from the fixed size of the coefficients of
$\Psi$ when it is decomposed into the eigenvectors of $\rho$, rather
than from the probability with which these eigenvectors are chosen.
This measure, too, is stationary under any unitary evolution that
leaves $\rho$ invariant. In particular, it is stationary in the
thermal case $\rho = \rho_{\Hilbert, H, \beta}$, and for generic $H$
it is an extremal stationary measure as characterized in
Section~\ref{sec:stationarity}; in fact it is, in the notation of the
last paragraph of Section~\ref{sec:stationarity}, $\lambda_p$ with
$p(E_n) = (1/Z) \exp(-\beta E_n)$.

This measure, too, is highly concentrated: For a Hilbert
space $\Hilbert$ of finite dimension $k$, it is supported by a
submanifold of real dimension $2k-m$ where $m$ is the number of
distinct eigenvalues of $H$, hence generically it is supported by a
submanifold of just half the dimension of $\Hilbert$.

\subsubsection{The Distribution of Guerra and Loffredo}

In \cite{GL81}, Guerra and Loffredo consider the canonical density
matrix $\rho_\beta$ for the one-dimensional harmonic oscillator and want to associate
with it a diffusion process on the real line, using stochastic
mechanics \cite{Nel85,Gol87}. Since stochastic mechanics associates a
process with every wave function, they achieve this by finding a
measure $\mu_\beta$ on $\sphere(L^2(\RRR))$ whose density matrix is
$\rho_\beta$.

They propose the following measure $\mu_\beta$, supported by coherent
states. With every point $(q,p)$ in the classical phase space $\RRR^2$
of the harmonic oscillator there is associated a coherent state
\begin{equation}\label{coherent}
  \psi_{q,p} (x) = (2\pi\sigma^2)^{-1/4} \, \exp \biggl( - \frac{(x-q)^2}
  {4\sigma^2} + \frac{i}{\hbar} xp - \frac{i}{2\hbar} pq \biggr)
\end{equation}
with $\sigma^2 = \hbar/2m\omega$, thus defining a mapping $C : \RRR^2
\to \sphere(L^2(\RRR))$,  $C(q,p) = \psi_{q,p}$. Let $H(q,p) =
p^2/2m + \tfrac{1}{2}m \omega^2 q^2$ be the classical Hamiltonian
function, and consider the classical canonical distribution at inverse
temperature $\beta'$,
\begin{equation}\label{rhocl}
  \rho^\mathrm{class}_{\beta'} (dq \times dp) = \frac{1}{Z'}
  e^{-\beta' H(q,p)} \, dq \, dp\,, \quad Z' = \int_{\RRR^2} dq\, dp\,
  e^{-\beta' H(q,p)}\,.
\end{equation}
Let $\beta' = \frac{e^{\beta \hbar \omega} -1}{\hbar \omega} \,.$ Then
$\mu_\beta = C_* \rho^\mathrm{class}_{\beta'}$ is the distribution on
coherent states arising from $ \rho^\mathrm{class}_{\beta'}$.
The density matrix of $\mu_\beta$ is $\rho_\beta$ \cite{GL81}.

This measure is concentrated on a 2-dimensional submanifold of
$\sphere(L^2(\RRR))$, namely on the set of coherent states (the image
of $C$). Note also that not every density matrix $\rho$ on $L^2(\RRR)$
can arise as the density matrix of a distribution on the set of
coherent states; for example, a pure state $\rho = |\psi \rangle
\langle \psi|$ can arise in this way if and only if $\psi$ is a
coherent state.

\subsubsection{The Distribution Maximizing an Entropy Functional}

In a similar spirit, one may consider, on a finite-dimensional Hilbert
space $\Hilbert$, the distribution $\gamma(d\psi) = f(\psi) \,
u_{\sphere (\Hilbert)} (d\psi)$ that maximizes the Gibbs entropy
functional
\begin{equation}\label{Gibbs}
  \mathscr{G}[f] = - \int\limits_{\sphere (\Hilbert)} \!\!\! u(d\psi)
  \:\: f(\psi) \, \log f(\psi)
\end{equation}
under the constraints that $\gamma$ be a probability distribution with
mean 0 and covariance $\rho_{\Hilbert, H, \beta}$:
\begin{subequations}
\begin{align}
  f &\geq 0\\
  \int\limits_{\sphere (\Hilbert)} \!\!\! u(d\psi) \, f(\psi) &= 1
  \label{fnorm}\\
  \int\limits_{\sphere (\Hilbert)} \!\!\! u(d\psi) \, f(\psi) \,
  | \psi \rangle &= 0\\
  \int\limits_{\sphere (\Hilbert)} \!\!\! u(d\psi) \, f(\psi) \, |\psi
  \rangle \langle \psi | &= \rho_{\Hilbert, H, \beta}\,. \label{fcov}
\end{align}
\end{subequations}
A standard calculation using Lagrange multipliers leads to 
\begin{equation}\label{fform}
  f(\psi) = \exp \langle \psi | L | \psi \rangle
\end{equation}
with $L$ a self-adjoint matrix determined by \eqref{fnorm} and
\eqref{fcov}; comparison with \eqref{mupowerlaw} shows that $\gamma$
is not a GAP measure.  (We remark, however, that another Gibbs entropy
functional, $\mathscr{G}'[f] = -\int_\Hilbert \lambda(d\psi) \:
f(\psi) \, \log f(\psi)$, based on the Lebesgue measure $\lambda$ on
$\Hilbert$ instead of $u_{\sphere (\Hilbert)}$, is maximized, under
the constraints that the mean be 0 and the covariance be $\rho$, by
the Gaussian measure, $f(\psi) \, \lambda(d\psi) = \G{\rho} (d\psi)$.)
There is no apparent reason why the family of $\gamma$ measures should
be hereditary.  

The situation is different for the microcanonical ensemble: here, the
distribution $u_{E, \delta} = \GAP{\rho_{E,\delta}}$ that we propose
is in fact the maximizer of the appropriate Gibbs entropy functional
$\mathscr{G}''$. Which functional is that? Since any measure $\gamma
(d\psi)$ on $\sphere(\Hilbert)$ whose covariance matrix is the
projection $\rho_{E,\delta} = \mathrm{const.} \, 1_{[E,E+\delta]}(H)$
must be concentrated on the subspace $\Hilbert_{E,\delta}$ and thus
cannot be absolutely continuous (possess a density) relative to
$u_{\sphere (\Hilbert)}$, we consider instead its density relative to
$u_{\sphere (\Hilbert_{E, \delta})} = u_{E, \delta}$, that is, we
consider $\gamma (d\psi) = f(\psi) \, u_{E, \delta} (d\psi)$ and set
\begin{equation}
  \mathscr{G}'' [f] = - \int\limits_{\sphere (\Hilbert_{E, \delta})}
  u_{E, \delta} (d\psi) \, f(\psi) \, \log f(\psi) \,.  
\end{equation} 
Under the constraints that the probability measure $\gamma$ have mean
0 and covariance $\rho_{E, \delta}$, $\mathscr{G}'' [f]$ is maximized
by $f \equiv 1$, or $\gamma = u_{E, \delta}$; in fact even without the
constraints on $\gamma$, $\mathscr{G}'' [f]$ is maximized by $f \equiv
1$.

\subsubsection{The Distribution of Brody and Hughston}
\label{sec:Brody}

Brody and Hughston \cite{Brody} have proposed the following distribution
$\mu$ to describe thermal equilibrium. They observe that the projective
space arising from a finite-dimensional Hilbert space, endowed with the
dynamics arising from the unitary dynamics on Hilbert space, can be
regarded as a classical Hamiltonian system with Hamiltonian function
$H(\CCC \psi) = \langle \psi| H | \psi \rangle / \langle \psi | \psi
\rangle$ (and symplectic form arising from the Hilbert space structure).
They then define $\mu$ to be the classical canonical distribution of this
Hamiltonian system, i.e., to have density proportional to $\exp(-\beta
H(\CCC \psi))$ relative to the uniform volume measure on the projective
space (which can be obtained from the symplectic form or, alternatively,
from $u_{\sphere(\Hilbert)}$ by projection from the sphere to the
projective space).  However, this distribution leads to a
density matrix, different from the usual one $\rho_\beta$ given by
\eqref{rhobetaH}, that does not describe the canonical ensemble.

\subsection{A Uniqueness Result for $\GAP{\rho}$}
\label{sec:uniqueness}

As $\EIG{\rho}$ is a family of measures \z{satisfying 
Properties~1--3} for \emph{most} density matrices $\rho$, the
question arises whether there is any family of measures, besides
$\GAP{\rho}$, satisfying these properties for \emph{all} density
matrices. We expect that the answer is no, and formulate the following
uniqueness conjecture: \textit{Given, for every Hilbert space
  $\Hilbert$ and every density matrix $\rho$ on $\Hilbert$, a
  probability measure $\mu(\rho)$ on $\sphere(\Hilbert)$ such that
  Properties~1--3 remain true when $\GAP{\rho}$ is
  replaced by $\mu(\rho)$, then $\mu(\rho) = \GAP{\rho}$.}  In other
words, we conjecture that $\mu = \GAP{\rho}$ is the only hereditary
covariant inverse of \eqref{rhomupsi}.

This is in fact true when we assume in addition that the mapping $\mu:
\rho \mapsto \mu(\rho)$ is suitably continuous. Here is the argument: When
$\rho$ is a multiple of a projection, $\rho = (\dim \Hilbert')^{-1}
P_{\Hilbert'}$ for a subspace $\Hilbert' \subseteq \Hilbert$, then
$\mu(\rho)$ must be, by covariance $U_* \mu(\rho) = \mu(U\rho U^*)$, the
uniform distribution on $\sphere(\Hilbert')$, and thus $\mu(\rho) =
\GAP{\rho}$ in this case. Consider now a composite of a system (system 1)
and a large heat bath (system 2) with Hilbert space $\Hilbert = \Hilbert_1
\otimes
\Hilbert_2$ and Hamiltonian $H = H_1 \otimes I_2 + I_1 \otimes H_2$, and
consider the 
microcanonical density matrix $\rho_{E,\delta}$ for this system. By
equivalence of ensembles, we have for suitable $\beta>0$ that
$\rho_{E,\delta}
\approx \rho_{\Hilbert, H, \beta} = \rho^{(1)}_\beta \otimes
\rho^{(2)}_\beta$ where $\rho^{(i)}_\beta = \rho_{\Hilbert_i, H_i,
  \beta}$. By the continuity of $\mu$ and $GAP$,
\[
  \mu \bigl( \rho^{(1)}_\beta \otimes \rho^{(2)}_\beta \bigr) \approx 
  \mu(\rho_{E,\delta}) = \GAP{\rho_{E,\delta}} \approx \GAP{
  \rho^{(1)}_\beta \otimes \rho^{(2)}_\beta}\,.
\]
Now consider, for a wave function $\Psi$ with distribution $\mu \bigl(
\rho^{(1)}_\beta \otimes \rho^{(2)}_\beta \bigr)$ respectively $\GAP{
  \rho^{(1)}_\beta \otimes \rho^{(2)}_\beta}$, the distribution of the
conditional wave function $\Psi_1$: by heredity, this is
$\mu(\rho^{(1)}_\beta)$ respectively $\GAP{\rho^{(1)}_\beta}$. Since
the distribution of $\Psi_1$ is a continuous function of the
distribution of $\Psi$, we thus have that $\mu(\rho^{(1)}_\beta) \approx
\GAP{\rho^{(1)}_\beta}$. Since we can make the degree of approximation
arbitrarily good by making the heat bath sufficiently large, we must have that
$\mu(\rho^{(1)}_\beta) = \GAP{\rho^{(1)}_\beta}$. For any density
matrix $\rho$ on $\Hilbert_1$ that does not have zero among its
eigenvalues, there is an $H_1$ such that $\rho = \rho^{(1)}_\beta =
Z^{-1} \, \exp(-\beta H_1)$ for $\beta =1$, and thus we have that
$\mu(\rho) = \GAP{\rho}$ for such a $\rho$; since these are dense, we have
that  $\mu(\rho) = \GAP{\rho}$ for all density matrices $\rho$ on
$\Hilbert_1$. Since $\Hilbert_1$ is arbitrary we are done.

\subsection{Dynamics of the Conditional Wave Function}

Markov processes in Hilbert space have long been considered (see
\cite{BP02} for an overview), particularly diffusion processes and
piecewise deterministic (jump) processes. This is often done for the
purpose of numerical simulation of a master equation for the density
matrix, or as a model of continuous measurement or of spontaneous wave
function collapse. Such processes could arise as follows.

Since the conditional wave function $\Psi_1$ arises from the wave
function $\sp{q_2}{\psi}$ by inserting a random coordinate $Q_2$ for
the second variable (and normalizing), any dynamics (i.e., time
evolution) for $Q_2$, described by a curve $t \mapsto Q_2(t)$ and
preserving the quantum probability distribution of $Q_2$, for example, as
given by  Bohmian mechanics \cite{DGZ}, gives rise to a
dynamics for the conditional wave function, $t \mapsto \Psi_1(t) =
\mathcal{N}(t) \, \sp{Q_2(t)}{\psi(t)}$, where $\psi(t)$
evolves according to Schr\"odinger's equation and $\mathcal{N}(t)=
\|\sp{Q_2(t)}{\psi(t)}
\|^{-1}$ is the normalizing factor. In this way one obtains a
stochastic process (a random path) in $\sphere(\Hilbert_1)$.  In the case
considered in Section~\ref{sec:hb1}, in which $\Hilbert_2$ corresponds to a
large heat bath, this process must have $\GAP{\rho_{\Hilbert_1,
    H_1,\beta}}$ as an invariant measure. It would be interesting to know
whether this process is approximately a simple process in
$\sphere(\Hilbert_1)$, perhaps a diffusion process, perhaps one of the
Markov processes on Hilbert space considered already in the literature.

\section{The Two-Level System as a Simple Example}
\label{sec:two}

In this last section, we consider a two-level system, with $\Hilbert =
\CCC^2$ and
\begin{equation}\label{H2}
  H = E_1 |1\rangle \langle 1| + E_2 |2\rangle \langle 2|,
\end{equation}
and calculate the joint distribution of the energy coefficients $Z_1 =
\sp{1}{\Psi}$ and $Z_2 = \sp{2}{\Psi}$ for a $\GAP{\rho
  _\beta}$-distributed $\Psi$ as explicitly as possible.  We begin
with a general finite-dimensional system, $\Hilbert = \CCC^k$, and
specialize to $k=2$ later.

One way of describing the distribution of $\Psi$ is to give its
density relative to the hypersurface area measure $u$ on
$\sphere(\CCC^k)$; this we did in \eqref{mudensity}. Another way of
describing the joint distribution of the $Z_n$ is to describe the
joint distribution of their moduli $|Z_n|$, or of $|Z_n|^2$, as the
phases of the $Z_n$ are independent (of each other and of the moduli)
and uniformly distributed, see \eqref{independ}.

Before we determine the distribution of $|Z_n|^2$, we repeat that its
expectation can be computed easily. In fact, for any $\phi \in
\Hilbert$ we have
\[
  \EEE \bigl| \sp{\phi}{\Psi} \bigr|^2 =
  \int\limits_{\sphere(\Hilbert)} \!\!  \GAP{\rho_\beta} (d\psi) \,
  \bigl| \sp{\phi}{\psi} \bigr|^2 \stackrel{\eqref{rhomupsi}}{=}
  \sp{\phi}{\rho_\beta | \phi} \stackrel{\eqref{rhobetaH}}{=}
  \frac{1}{Z(\beta)} \sp{\phi}{e^{-\beta H} | \phi}.
\]
Thus, for $|\phi\rangle = |n\rangle$, we obtain $\EEE |Z_n|^2 =
e^{-\beta E_n}/\tr\, e^{-\beta H}$.

For greater clarity, from now on we write $Z_n^{GAP}$ instead of
$Z_n$.  A relation similar to that between $\GAP{\rho}$, $\GA{\rho}$,
and $\G{\rho}$ holds between the joint distributions of the
$|Z_n^{GAP}|^2$, of the $|Z_n^{GA}|^2$, and of the $|Z_n^G|^2$. The
joint distribution of the $|Z_n^G|^2$ is very simple: they are
independent and exponentially distributed with means $p_n = e^{-\beta
  E_n} / Z(\beta)$. Since the density of $GA$ relative to $G$, $dGA/dG
= \sum_n |z_n|^2$, is a function of the moduli alone, and since,
according to \eqref{independ}, $GA = GA_\mathrm{phases} \times
GA_\mathrm{moduli}$, we have that
\[
  GA_\mathrm{moduli} = \sum_n |z_n|^2 \,
  G_\mathrm{moduli}.
\]
Thus,
\begin{equation}\label{jointZN}
  \PPP \bigl( |Z_1^{GA}|^2 \in ds_1 , \ldots, |Z_k^{GA}|^2 \in ds_k \bigr)
  = \frac{s_1+ \ldots + s_k} {p_1 \cdots p_k} \exp \Bigl(
  -\sum_{n=1}^k \frac{s_n}{p_n} \Bigr) ds_1 \cdots ds_k,
\end{equation}
where each $s_n \in (0,\infty)$. Finally, the $|Z_n^{GAP}|^2$ arise by
normalization,
\begin{equation}\label{ZZN} 
  |Z_n^{GAP}|^2 = \frac{|Z_n^{GA}|^2}{\sum\limits_{n'}
   |Z_{n'}^{GA}|^2}.
\end{equation}

We now specialize to the two-level system, $k=2$. Since $|Z_1^{GAP}|^2
+|Z_2^{GAP}|^2 =1$, it suffices to determine the distribution of
$|Z_1^{GAP}|^2$, for which we give an explicit formula in
\eqref{Z12dist} below.  We want to obtain the marginal distribution of
\eqref{ZZN} from the joint distribution of the $|Z_n^{GA}|^2$ in
$(0,\infty)^2$, the first quadrant of the plane, as given by
\eqref{jointZN}. To this end, we introduce new coordinates in the
first quadrant:
\[
  s=\frac{s_1}{s_1 + s_2}, \quad \lambda = s_1 +s_2,
\]
where $\lambda>0$ and $0<s<1$.  Conversely, we have $s_1 = s\lambda$
and $s_2 = (1-s) \lambda$, and the area element transforms according
to
\[
  ds_1 \, ds_2 = \Bigl| \det \frac{\partial (s_1,s_2)}{\partial
  (s,\lambda)} \Bigr| ds \, d\lambda = \lambda \, ds \, d\lambda.
\]

\begin{figure}[h]
\begin{center}
\includegraphics[width=.7 \textwidth]{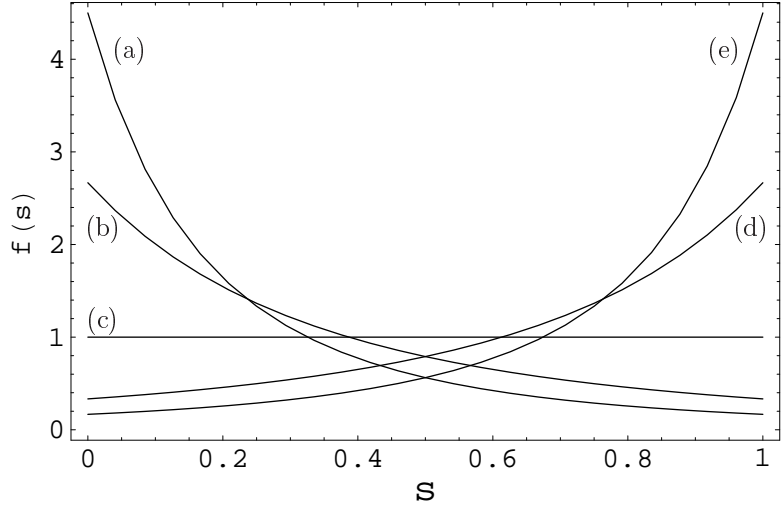}
\end{center}
\caption{Plot of the distribution density function $f(s)$ of
  $|Z_1|^2$, defined in \eqref{Z12dist}, for various values of the
  parameter $\delta = \exp(\beta(E_2-E_1))$: (a)~$\delta = 1/3$,
  (b)~$\delta = 1/2$, (c)~$\delta = 1$, (d)~$\delta = 2$, (e)~$\delta
  = 3$.}
\label{figtwo}
\end{figure}

Therefore, using
\[
  \int\limits_0^\infty d\lambda \, \lambda^2 e^{- x\lambda}= 2 x^{-3}
  \text{ for }x>0,
\]
we obtain
\begin{subequations}
\begin{align}
  \PPP \Bigl( |Z_1^{GAP}|^2 \in ds \Bigr) &= ds \int\limits_0^\infty d
  \lambda \, \frac{e^{\beta \tr\, H}}{Z(\beta)} \lambda^2 \, \exp \Bigl(
  -\lambda \bigl( e^{\beta E_1} s + e^{\beta E_2} (1-s) \bigr) \Bigr)
  =\\ &= \frac{2e^{\beta \tr\, H}}{Z(\beta)} \bigl( e^{\beta E_1} s +
  e^{\beta E_2} (1-s) \bigr)^{-3} ds = \\
  &= \bigl( \alpha_1 s + \alpha_2 (1-s) \bigr)^{-3} ds \label{Z12dist}
  =: f(s) \, ds\,, \quad 0<s<1,
\end{align}
\end{subequations}
with $\alpha_1 = (\delta^{-1} (\delta^{-1} +1)/2)^{1/3}$ and $\alpha_2
= (\delta (\delta +1)/2)^{1/3}$ for $\delta = \exp(\beta(E_2-E_1))$.
The density $f$ of the distribution \eqref{Z12dist} of $|Z_1^{GAP}|^2$
is depicted in Figure~1 for various values of $\delta$. For
$\delta=1$, $f$ is identically 1. For $\delta >1$, we have $\alpha_2 =
\delta \alpha_1 > \alpha_1$, so that $\alpha_1 s + \alpha_2 (1-s)$ is
decreasing monotonically from $\alpha_2$ at $s=0$ to $\alpha_1$ at
$s=1$; hence, $f$ is increasing monotonically from $\alpha_2^{-3}$ to
$\alpha_1^{-3}$. For $\delta<1$, we have $\alpha_2< \alpha_1$, and
hence $f$ is decreasing monotonically from $\alpha_2^{-3}$ to
$\alpha_1^{-3}$. In all cases $f$ is convex since $f'' \geq 0$.

\bigskip

\noindent \textit{Acknowledgments.} We thank Andrea Viale (Universit\`a di
Genova, Italy) for preparing the figure, Eugene Speer (Rutgers University,
USA) for comments on an earlier version, James Hartle (UC Santa Barbara,
USA) and Hal Tasaki (Gakushuin University, Tokyo, Japan)
for helpful comments and suggestions, Eric Carlen (Georgia Institute
of Technology, USA), Detlef D\"urr (LMU M\"unchen, Germany), Raffaele
Esposito (Universit\`a di L'Aquila, Italy), Rossana Marra (Universit\`a di
Roma ``Tor Vergata'', Italy), and Herbert Spohn (TU M\"unchen, Germany) for
suggesting references, \zz{and Juan Diego Urbina (the Weizmann Institute,
Rehovot, Israel) for bringing the connection between Gaussian random wave
models and quantum chaos to our attention.}  We are grateful for the
hospitality of the Institut des Hautes \'Etudes Scientifiques
(Bures-sur-Yvette, France), where part of the work on this paper was done.

\zz{The work of S.~Goldstein was supported in part by NSF Grant
DMS-0504504, and that of} J.~Lebowitz by NSF Grant DMR 01-279-26 and AFOSR
Grant AF 49620-01-1-0154. The work of R.~Tumulka was supported by INFN and
by the European Commission through its 6th Framework Programme
``Structuring the European Research Area'' and the contract
Nr. RITA-CT-2004-505493 for the provision of Transnational Access
implemented as Specific Support Action. The work of N.~Zangh\`\i\ was
supported by INFN.

\end{document}